\documentstyle{mn}
\input psfig

\newcount\eqnno \eqnno=0 
\def\numeqn{\global\advance\eqnno by 1 \eqno(\the\eqnno)}

\title[An excursion set model]
{An excursion set model for the distribution of dark matter and 
dark matter haloes}

\author[R. K. Sheth]
{Ravi K. Sheth\\
Max-Planck Institut f\"ur Astrophysik, Karl-Schwarzschild-Str. 1,
85740 Garching, Germany\\
\smallskip
Email: sheth@mpa-garching.mpg.de
}

\date{Submitted 1998 April; in original form 1997 September}

\begin{document}

\maketitle

\begin{abstract}
A model of the gravitationally evolved dark matter distribution, 
in the Eulerian space, is developed.  It is a simple extension of 
the excursion set model that is commonly used to estimate the 
mass function of collapsed dark matter haloes.  In addition to 
describing the evolution of the dark matter itself, the model 
allows one to describe the evolution of the Eulerian space 
distribution of the haloes.  It can also be used to describe 
density profiles, on scales larger than the virial radius, of 
these haloes, and to quantify the way in which matter flows in 
and out of Eulerian cells.  When the initial Lagrangian space 
distribution is white noise Gaussian, the model suggests that the 
Inverse Gaussian distribution should provide a reasonably good 
approximation to the evolved Eulerian density field, in agreement 
with numerical simulations.  Application of this model to 
clustering from more general Gaussian initial conditions is 
discussed at the end.  

\end{abstract} 
\maketitle

\section{Introduction}
The hypothesis that, in comoving coordinates, initially denser
regions contract more rapidly than less dense regions, and that 
sufficiently underdense regions expand, is simple, reasonable, 
and powerful.  As a consequence of this expansion and contraction, 
the density distribution in the initial Lagrangian space will be 
different from that in the evolved, Eulerian space.  
Suppose that, as the universe evolves, the number of 
expanding and contracting regions is conserved---only their 
comoving size changes---and the mass within each such region 
is also conserved.  If we have a model for the way in which 
the evolution of the size of a region depends on its initial size 
and density, and we also have a model for the initial number of 
regions as a function of initial size and density, then we can 
compute the distribution of sizes and densities at some later 
time.  For example, suppose that the initial Lagrangian density 
distribution is a Gaussian random field, and that the evolution 
of regions is given by the spherical collapse model.  Then it 
should be possible to construct a model for $p(M_0|R,z)$, where 
$p(M_0|R,z)$ is the fraction of regions of size $R$ that, at $z$, 
contain mass $M_0$.   The quantity $p(M_0|R,z)$ is often called 
the Eulerian probability distribution function.  

In the Press-Schechter (1974) approach, at any given time, all 
matter is in the form of collapsed objects, usually called haloes, 
and the distribution of halo masses evolves with time.  At any time, 
the matter within a randomly placed cell $R$ is divided among many 
collapsed haloes.  Thus, $p(M_0|R,z)$ depends both on the halo mass 
function, and on the spatial distribution of the haloes.  
Bond et al. (1991) showed how to estimate the evolution of the 
halo mass function if the initial Lagrangian distribution is 
Gaussian (also see Lacey \& Cole 1993).  
They did not show how to estimate the spatial distribution 
of these haloes, but Mo \& White (1996) showed how this might be 
accomplished.  This paper develops a model that combines, 
self-consistently, the Bond et al. (1991) excursion set approach 
with the Mo \& White (1996) model for the Eulerian space halo 
distribution.  The model developed here allows one to 
simultaneously describe both the distribution of the dark matter, 
i.e. $p(M_0|R,z)$, and that of the haloes.  
See, e.g., Mo \& White (1996) for why such a model is useful.  

Section~\ref{model} describes the model.  
It shows why the distribution of first crossings of a barrier whose 
height is not constant, by Brownian motion random walks, is useful.   
The shape of the barrier associated with the spherical collapse model 
is given in Section~\ref{smodel}.   
The relation between the first crossing distribution and the Eulerian 
distribution $p(M_0|R,z)$ is discussed in Section~\ref{prelim}.  
Section~\ref{halod} shows that, in the context of the model 
developed here, the halo mass function is related to the small cell 
size limit (i.e. $R\to 0$) of $p(M_0|R,z)$.  
It discusses the Bond et al. (1991) excursion set results in 
this context, and then shows how the model can be used to describe 
the spatial distribution of the haloes as well.  The results of 
Mo \& White (1996) are discussed in Section~\ref{mowhite}.  
Section~\ref{twobare} shows that the associated two barrier problem 
can be used to provide information about the evolved density 
profile, and also about the way in which matter flows in and out 
of Eulerian cells.  

The first crossing distribution associated with the 
spherical collapse barrier must be obtained numerically.  
Therefore, to illustrate the usefulness of our approach, 
Section~\ref{linear} shows the results of assuming that the 
initial distribution is white-noise Gaussian, and that the 
barrier shape is simpler than that associated with the spherical 
collapse model.  In Section~\ref{linear}, the barrier is assumed 
to be linear for a number of reasons.  
Firstly, this linear barrier can be understood as arising from a 
simple variant of the spherical collapse model (Section~\ref{scoll}). 
Secondly, the barrier crossing distribution can be computed 
analytically (e.g. Schr\"odinger 1915).  The details of the 
derivation are presented in Appendix~\ref{deriv}.  
Thirdly, the Eulerian probability distribution associated with the 
first crossings of this linear barrier is Inverse Gaussian 
(Section~\ref{pdfe}), and the Inverse Gaussian provides a good fit 
to the Eulerian distribution measured in numerical simulations of 
clustering from white noise initial conditions (Section~\ref{sims}).  
Finally, studies of clustering from Poisson initial conditions also 
suggest that this barrier shape is a useful approximation (Sheth 1998).  

The Bond et al. (1991) results are derived, within the context 
of this linear barrier model, in Section~\ref{const}.  
Analytic expressions for the Eulerian distribution function as well 
as the halo--mass and halo--halo correlations are derived for all 
scales, and for all times, in Sections~\ref{pdfe} and~\ref{halobar}.  
Section~\ref{twobar} provides various formulae associated with 
the two-linear-barrier problem.  
A final section summarizes the results.  It also discusses 
how the model can be extended to describe clustering from 
more general initial conditions.  

\section{The excursion set model}\label{model}
The first subsection defines various Eulerian space quantities 
of interest in this paper.  Since these definitions are standard
(e.g. Peebles 1980), no further references are given.  
Subsequent subsections develop a model which allows one to estimate 
the evolution of these quantities.  

\subsection{The matter and halo distribution in Eulerian space}\label{defs}
Imagine partitioning the Eulerian space $V_{\rm tot}$ into 
a large number of cells, each of size $V\equiv (4\pi/3)R^3$, at 
time $z$.  Since the volume of each cell is $V$, the total number 
of such cells is $N_{\rm tot}=V_{\rm tot}/V$.  
Let $p(M_0|R,z)\,{\rm d}M_0$ denote the fraction of these cells 
that contain mass between $M_0$ and $M_0+{\rm d}M_0$.  Then, 
at a given $z$, 
\begin{equation}
p(M_0|R,z)\,{\rm d}M_0 = {N(M_0|R)\,{\rm d}M_0\over N_{\rm tot}}
= {V\,N(M_0|R)\,{\rm d}M_0\over V_{\rm tot}} ,
\label{pmr}
\end{equation}
where we have not bothered to write $z$ explicitly on the right 
hand side.  
Define $\Delta\equiv (1+\delta)\equiv M_0/\bar\rho V$, where 
$\bar\rho$ is the average comoving density in the Eulerian space.  
That is,
\begin{equation}
\bar\rho = {M_{\rm tot}\over V_{\rm tot}} = {1\over V_{\rm tot}}
\int_0^\infty M_0\,N(M_0|R,z)\,{\rm d}M_0.  
\end{equation}
Then 
\begin{equation}
p(\Delta|R,z)\,{\rm d}\Delta = p(M_0|R,z)\,{\rm d}M_0 ,
\label{pdr}
\end{equation}
and $p(\Delta|R,z)$ is the probability distribution function 
of the density in Eulerian space at $z$.  Since 
${\rm d}\Delta = {\rm d}M_0/(\bar\rho V)$, 
\begin{equation}
\int_0^\infty \!\!p(\Delta|R,z)\ {\rm d}\Delta = 
\int_0^\infty \!\!\Delta\,p(\Delta|R,z)\ {\rm d}\Delta = 1
\label{norm}
\end{equation}
for all $z$.  

Let $\bar n(M_1,\delta_{\rm c1})$, where 
$\delta_{\rm c1} = \delta_{\rm c0}(1+z_1)$ and $\delta_{\rm c0}$ 
is some constant that will be determined later, denote the average 
number density of $M_1$ haloes identified at $z_1$.  
On average, the number of such haloes within an Eulerian cell $V$ 
is $\bar n(M_1,\delta_{\rm c1})V$.  
Let $N_1$ denote the number of such $M_1$ haloes within the 
$i$th Eulerian cell.  Suppose we classify all Eulerian cells by 
the mass $M_0$ within them at some $z<z_1$.  
Let $N(1|0)$ denote $\langle N_1\rangle_0$, where the average 
is over only those Eulerian cells that contain mass $M_0$ at $z$.  
Clearly, 
\begin{equation}
\bar n(M_1,\delta_{\rm c1})V = 
\int_0^\infty N(1|0)\ p(\Delta|R,z)\,{\rm d}\Delta ,
\label{nme}
\end{equation}
since equation~(\ref{pdr}) shows that 
$p(M_0)\,{\rm d}M_0 = p(\Delta)\,{\rm d}\Delta$.

Define 
\begin{equation}
\delta_{\rm h}(1|0) = 
{N(1|0)\over \bar n(M_1,\delta_{\rm c1})V} - 1.  
\end{equation}
This is the number, relative to the average number of 
$(M_1,\delta_{\rm c1})$ haloes in Eulerian cells $V$, of such 
haloes that are within Eulerian cells which contain mass $M_0$ 
at $z$, minus one.  
The cross correlation $\bar\xi_{\rm hm}(M_1,\delta_{\rm c1}|R,z)$ 
between haloes and mass in Eulerian space, averaged over 
Eulerian cells of comoving size $V$, at the epoch $z$, is 
\begin{eqnarray}
\bar\xi_{\rm hm}(1|R,z) &=& 
\Bigl\langle \delta_{\rm h}(1|0)\ \delta \Bigr\rangle_{\rm R} \nonumber \\
&=& \int_0^\infty \!\!\!(\Delta-1)\,
{N(1|0)\,p(\Delta|R,z)\over \bar n(M_1,\delta_{\rm c1})V}
\ {\rm d}\Delta ,
\label{xihme}
\end{eqnarray}
where we have used equation~(\ref{norm}) to set  
$\langle\Delta\rangle = \langle 1+\delta\rangle = 1$, and so 
$\langle\delta\rangle=0$.  

Let $\bar\xi_{\rm hh}(M_1,M_2,\delta_{\rm c1}|R,z)$ denote the 
(volume average of the) correlation function of $M_1$ and $M_2$ 
haloes identified at the epoch $z_1$, averaged over all comoving 
Eulerian cells $V$, at the epoch $z\le z_1$.  
This average can be computed in two steps.  Let $N_1$ and $N_2$ 
denote the number of $M_1$ and $M_2$ haloes within an Eulerian 
cell $V$.  Classify all Eulerian cells by the mass $M_0$ within 
them at $z$.  Let $C(12|0)$ denote the average over all $M_0$ 
cells of $(N_1N_2)$:  $C(12|0)\equiv\langle N_1N_2\rangle_0$.  Then 
\begin{equation}
1+\bar\xi_{\rm hh}(12|R,z)\!=\! 
\int_0^\infty\!\!\!\!\!\!
{C(12|0)\ p(\Delta|R,z)\over 
\bar n(M_1,\delta_{\rm c1})V\,\bar n(M_2,\delta_{\rm c1})V}\ 
{\rm d}\Delta .
\label{xihhe}
\end{equation}
Higher order moments of the halo distribution can be defined 
similarly.  The remainder of this section develops a 
barrier-crossing, excursion set model which allows one to 
estimate all these Eulerian quantities.  

\subsection{The spherical collapse model}\label{smodel}
We will assume that the total mass and comoving volume of the 
evolved Eulerian space is the same as that initially (i.e., 
in the Lagrangian space).  Then the average density in the two 
spaces is the same:  $\bar\rho_0=\bar\rho$, where here and below 
quantities with subscript zero are in the Lagrangian space.  

We will also assume that the initial overdensity fluctuations 
in the Lagrangian space are small:  $\delta_0\ll 1$ initially.  
If $\bar\rho$ is the average background density, then 
$M_0 = \bar\rho_0 V_0(1+\delta_0)\approx\bar\rho V_0$, where 
$V_0 = (4\pi/3)\,R_0^3$.  That is, the initial mass $M_0$, volume 
$V_0$ and size $R_0$ are all equivalent variables.  Consider a 
region that initially contains mass $M_0$ and has initial 
overdensity $\delta_0$.  
At some later time, it has size $R(z)$, so that the overdensity 
within it is $\delta$, where 
\begin{equation}
1 + \delta \equiv \Delta = M_0/\bar\rho V = (R_0/R)^3 .
\label{defdelta}
\end{equation}
Notice that $\Delta$ is the same Eulerian quantity of the 
previous subsection, and that it is simply the ratio of the 
Lagrangian volume to the Eulerian volume of the region.  

The spherical collapse model (e.g. Peebles 1980) allows one to 
describe the evolution of such a region.  
In particular, it provides relations between the initial size 
$R_0$, the initial overdensity $\delta_0$, the time $z$, and the 
evolved Eulerian size at that time $R(z)$.  
If $R_0$, $\delta_0$, and $z$ are given, then 
$R(R_0,\delta_0,z)$ is determined by the model.  
On the other hand, if only $R$ and $z$ are given, then the model 
describes a curve in the $(R_0,\delta_0)$ plane:  
$\delta_0(R_0|R,z)$.  To a good approximation, this relation is 
\begin{equation}
{\delta_0(R_0|R,z)\over 1+z} =
1.68647 - {1.35\over\Delta^{2/3}} - {1.12431\over\Delta^{1/2}}
+ {0.78785\over\Delta^{0.58661}}
\label{mow}
\end{equation}
(Mo \& White 1996).  
A simpler approximation to this relation is 
\begin{equation}
{\delta_0(R_0|R,z)\over 1+z} = 
\delta_{\rm c0} - \delta_{\rm c0}\,\Delta^{-1/\delta_{\rm c0}}
\label{bnrdo}
\end{equation}
(Bernardeau 1994).  Whereas Bernardeau used $\delta_{\rm c0}=1.5$, 
the value $1.68647$ is also acceptable.  

Notice that when $R\to 0$, then $\Delta\to\infty$, so 
formulae~(\ref{mow}) and~(\ref{bnrdo}) both become 
$\delta_0\to \delta_{\rm c0}\,(1+z)$ with $\delta_{\rm c0}=1.686$.  
In this limit $\delta_0(R_0|R=0,z)$ is independent of $R_0$.  
When $R>0$, then $\delta_0(R_0|R,z)$ decreases monotonically 
as $R_0$ decreases.  
Essentially, equation~(\ref{mow}) shows that a given pair $R$ 
and $z$ could initially have come from a range of $R_0$ and 
$\delta_0$.  This is sensible; in the 
spherical model, initially denser regions collapse more rapidly 
than less dense regions.  Therefore, a region of size $R$ at $z$ 
may initially have been a small region containing a small 
overdensity, or it may have been larger initially, but with a 
correspondingly larger overdensity.  

Our model for estimating the Eulerian probability distribution 
function $p(M_0|R,z)$ works as follows.  At any given $z$, imagine 
partitioning space into a large number of cells each of size 
$V$.  Assume that each cell evolved according to the 
spherical model independently of the others.  This means that 
the mass within each cell remains the same---only its comoving 
size changes---and the total number of cells is conserved.  
These are strong simplifying assumptions that do not have a 
rigorous physical justification.  
However, they allow one to estimate a 
number of useful quantities.  As with the Bond et al. (1991) 
excursion set approach, the extent to which the model here is able 
to reproduce the results of numerical simulations is the only real 
justification for these assumptions.  

Since each cell evolved according to the spherical model, and the 
number of such cells is conserved, to compute $p(M_0|R,z)$ we simply 
need to specify the relative numbers of regions initially with 
$(R_0,\delta_0)$ that have now evolved into regions $(R,z)$.  
Clearly, this distribution depends on the initial distribution 
of fluctuations and on the evolution described by the spherical model.  
The next subsection shows how to do this.  

\subsection{The first crossing distribution}\label{prelim}
Consider a density field $\delta_0({\bmath r};0)$.  Recall that 
the subscript zero denotes the fact that $\delta_0$ is a quantity 
measured in the Lagrangian space.  Smoothing this field with a 
window $W$ of scale $R_0$ produces a smoothed field 
$\delta_0({\bmath r};R_0)$:  
\begin{eqnarray}
\delta_0({\bmath r};R_0) &=& 
\int W(|{\bmath r}-{\bmath r}'|,R_0)\ \delta_0({\bmath r}';0)\ 
{\rm d}{\bmath r}' \nonumber \\
&=& \int \bar W(k,R_0)\ \delta_0(k)\,\exp(i{\bmath k.r})\ 
{\rm d}{\bmath k},
\end{eqnarray}
where $\bar W$ is the Fourier transform of the window $W$.  
By definition of the average density, $\delta_0\to 0$ as 
$R_0\to\infty$, for all positions ${\bmath r}$.  
Thus, for each position in the Lagrangian space there is a curve, 
$\delta_0(R_0)$, which describes the overdensity $\delta_0$ 
in a window of Lagrangian size $R_0$ centred on that position.  
Call such a curve a trajectory.  The volume associated with a 
window of size $R_0$ is $V_0\equiv V_{\rm W}R_0^3$, where 
$V_{\rm W}$ is some constant that may depend on the shape of the 
window, but does not depend on $R_0$.  For example, 
$V_{\rm W}=4\pi/3$ for a Top Hat window.  The mass 
within this volume is $M_0=\bar\rho_0V_0(1+\delta_0)$, where 
$\bar\rho_0$ is a constant that denotes the average mass density 
in the Lagrangian space.  Provided $\delta_0$ is small, 
$M_0\approx\bar\rho_0V_0$ to lowest order.  In this approximation, 
there is a deterministic relation between the volume $V_0$ 
and the mass within it:  $M_0\equiv \bar \rho_0 V_0$ in the 
Lagrangian space.  

There is also a deterministic relation between the scale $R_0$, 
and the variance $S_0$ associated with that scale:
\begin{equation}
S_0 = \bigl\langle \delta^2_0({\bmath r};R_0)\bigr\rangle = 
\int P(k)\,\bar W^2(k;R_0)\ {\rm d}^3k ,  
\label{slin}
\end{equation}
where $P(k)$ is the power spectrum of the unsmoothed Lagrangian 
field.  Thus, for a given $P(k)$, the quantities $S_0$, $R_0$, 
$V_0$ and $M_0$ are equivalent.  For most power spectra of 
interest, $S_0$ increases as $R_0$ decreases.  For what 
follows, assume that $S_0\to\infty$ as $R_0\to 0$.  

The deterministic relation between $R_0$ and $S_0$ means that 
to each trajectory $\delta_0(R_0)$ associated with a position 
in the Lagrangian space, there is a corresponding trajectory 
$\delta_0(S_0)$.  Since $\delta_0(R_0)\to 0$ as $R_0\to\infty$, 
all these trajectories start from the origin 
$(S_0,\delta_0) = (0,0)$.  Since $S_0\to\infty$ as $R_0\to 0$, 
and since $S_0$ is a measure of the mean square distance of 
a trajectory at $S_0$ from the $S_0$ axis, any trajectory may 
be an arbitrary distance above or below the $S_0$ axis as 
$R_0\to 0$.  

\begin{figure}
\centering
\mbox{\psfig{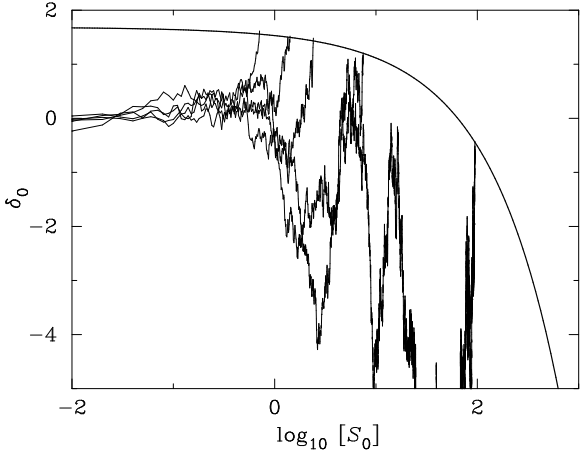}}
\caption{Examples of trajectories (thin jagged curves) traced 
out by the Lagrangian overdensity, $\delta_0$, as a function 
of linear variance, $S_0$.  The trajectories are absorbed at 
the barrier (thick solid line).  Here, the barrier shape is given 
by the spherical collapse model (equation~\ref{mow}), 
and $S_0\propto 1/V_0$ as it is for white noise.  }
\label{bmotion}
\end{figure}

The deterministic relation between $R_0$ and $S_0$ also means 
that equation~(\ref{mow}) represents a curve in the $(S_0,\delta_0)$ 
space.  Given $R$ and $z$, draw this curve, and call it the 
barrier $B(S_0|R,z)$.  The thick solid curve in Fig.~\ref{bmotion} 
shows this barrier for a representative value of $R$ and $z$.  
It decreases monotonically as $S_0$ increases.  
For a given $z$, equation~(\ref{mow}) shows that curves for different 
$R$ all have the same value $B=\delta_{\rm c0}(1+z)$ at 
$S_0=0$.  Equation~(\ref{mow}) also shows that $B=0$ when 
$R=R_0$, so curves for larger values of $R$ cross the $S_0$ axis at 
smaller values of $S_0$.  In fact, for a given $z$, the curves for 
different $R$ form a nested family (see Fig.~\ref{barrier}).  
This will be important below.  

Since $B(S_0|R,z)$ decreases monotonically with increasing $S_0$, 
each Lagrangian trajectory will intersect the barrier at least 
once (see Fig.~\ref{bmotion}).  Define the first crossing as the 
smallest value of $S_0$ at which the Lagrangian trajectory 
$\delta_0(S_0)$ intersects $B(S_0|R,z)$.  That is, it is 
the smallest $S_0$ at which $\delta_0=B$.  
Bond et al. (1991) discuss why the first crossing distribution 
is so important (it solves the so-called cloud-in-cloud problem).  
Although they only considered the special case in which 
$B(S_0|R,z)$ was independent of $S_0$, their argument 
holds for the more general barrier considered here also.  
The argument is as follows.  

Since $S_0$ and $M_0$ are equivalent variables, the first crossing 
value of $S_0$ represents the largest value of $M_0$ at which the 
trajectory had $\delta_0(S_0) = B$.  Since $M_0$ and $V_0$ are 
equivalent variables, the trajectory can be thought of as 
representing a volume element of a Lagrangian volume $V_0$ 
that has Eulerian size $R$ at $z$.  
Suppose we agree that the barrier crossings are significant, and 
the mass we associate with a trajectory could be that associated 
with any of the crossings.  If these possible assignments correspond 
to $M_0>M_1>\cdots$, then the trajectory represents concentric 
Lagrangian regions $V_0>V_1>\cdots$.  In effect, Bond et al. (1991) 
assume that successive shells never cross; whereas the actual sizes of 
$V_0>V_1>\cdots$ may change, the order of the shells is preserved.  
This means that if the largest region $V_0$ has Eulerian size $V$, 
then the subregions originally within it will remain within it.  
So the largest Lagrangian region associated with the trajectory, 
i.e., the one associated with the first crossing, is the one that 
is relevant.  Counting the largest region $V_0$ at once includes all 
the smaller ones within it, and is the natural way to avoid 
double-counting the smaller regions.  

Although Bond et al. (1991) only considered the case in which 
Lagrangian regions had collapsed to Eulerian size $R=0$, their 
argument can also be applied in the $R>0$ case studied here.  
Recall that, for a given $z$, curves for different $R$ are 
nested.  Consider a trajectory $\delta_0(S_0)$ that crosses the 
barrier $B(R)$ for the first time at $V_0$.  
The Lagrangian scale $V_{n-1}$, corresponding to the $n$th crossing 
of $B(R)$ by this same trajectory may be smaller than the Lagrangian 
scale $V_0'$ corresponding to the first crossing, by this trajectory, 
of the barrier $B(R')$ for some Eulerian $R'\le R$.  
In our model, such a trajectory represents a volume element of a 
Lagrangian region $V_0$ that at $z$, has Eulerian size $R$, at which 
time it contains a subregion that originally had size $V_0'\le V_0$ 
and now has size $R'\le R$.  The assumption is that if $V_0'$ has 
Eulerian size $R'$, and if $V_{n-1}<V_0'\le V_0$ originally, then 
all the mass $M_{n-1}$ contained within $V_{n-1}$ must still be 
contained within the Eulerian region $R'$ now occupied by $V_0'$.  
Thus, the model developed here explicitly preserves the ordering 
that was assumed by Bond et al. (1991), and so the first crossing 
of the barrier $B(S_0|R,z)$ remains the one that is significant.  

Let $f(S_0,\delta_0|R,z)\,{\rm d}S_0$ denote the fraction of 
trajectories that have their first crossing of the barrier between $
S_0$ and $S_0+{\rm d}S_0$.  Then, for a given barrier $B(S_0|R,z)$, 
$f(S_0,\delta_0|R,z)\,{\rm d}S_0$ can be equated with the fraction 
of Lagrangian space that is associated with regions containing 
mass $M_0$ that each occupy an Eulerian region $R$ at $z$.  
If $N(M_0|R,z)\,{\rm d}M_0$ denotes the number of such regions, 
then each such region occupied a Lagrangian volume 
$V_0=M_0/\bar\rho_0$, so 
\begin{equation}
f(S_0,\delta_0|R,z)\,{\rm d}S_0 = 
{V_0\,N(M_0|R,z)\,{\rm d}M_0\over V_{\rm tot}} ,
\label{fsnm}
\end{equation}
where $V_{\rm tot} \equiv \int V_0\,N(M_0|R,z)\,{\rm d}M_0$.  
Thus, the distribution of first crossings gives an estimate 
of the number density of Lagrangian regions which contained mass 
$M_0$ and had initial overdensity $\delta_0$, so that at $z$, 
while they still contain mass $M_0$, they occupy the Eulerian 
volume $V$.  

If we further suppose that the number of such regions is the same 
in both the Eulerian and Lagrangian spaces, then comparison 
with equation~(\ref{pmr}) shows that 
\begin{equation}
p(M_0|R,z)\,{\rm d}M_0 = 
{V\over V_0}\ f(S_0,\delta_0|R,z)\,{\rm d}S_0 .  
\label{conserve}
\end{equation}
Since $(V_0/V) = M_0/(\bar\rho V) = (1+\delta) = \Delta$, 
this means that 
\begin{equation}
f(S_0,\delta_0|R,z)\ {\rm d}S_0 = 
\Delta\,p(\Delta|R,z)\ {\rm d}\Delta,  
\label{fspd}
\end{equation} 
provided that $\delta_0$ is the function of $S_0$, $R$ and $z$ 
that is given by equation~(\ref{mow}).  
This shows how the barrier crossing distribution is related 
to $p(\Delta|R,z)$.  

Notice that the left hand side of equation~(\ref{fspd}) is 
determined by Lagrangian space quantities 
(the trajectories which cross the barrier are Lagrangian), 
whereas the right hand side is associated with Eulerian space.  
A similar sort of relation between Lagrangian and Eulerian 
space quantities was used by Bernardeau (1994) (see 
the discussion at the start of his Section 3.2.2).  The difference 
between his work and ours is that he used a Lagrangian distribution 
that was derived from perturbation theory, rather than from a 
barrier crossing model, for the left hand side of 
equation~(\ref{fspd}).  

In summary:  
The Eulerian probability distribution function $p(M_0|R,z)$ is 
the fraction of regions of size $R$ that, at $z$, contain mass 
$M_0$.  To estimate it, first assume that the initial fluctuation 
$\delta_0$ is small, so $M_0/\bar\rho_0 = V_0 = (4\pi/3)R_0^3$.  
Next, choose a random position in the initial field.  
Compute $\delta_0$ in spheres $R_0$ centred on this position.  
Call the curve $\delta_0(R_0)$  centred on this position a trajectory. 
Given $R$ and $z$, draw the barrier $B(R_0|R,z)$ associated with 
the spherical collapse model (equation~\ref{mow}).  
The trajectory $\delta_0(R_0)$ may intersect the barrier 
$B(R_0|R,z)$ many times.  
Find the largest value of $R_0$ at which the trajectory intersects 
$B(R_0|R,z)$.  Call this the first crossing of $B(R_0|R,z)$.  
Associate mass $M_0$ with this trajectory.  Since mass $M_0$ and 
initial volume $V_0$ are equivalent, this trajectory represents a 
volume element of a region containing mass $M_0$.  
Initially, $M_0$ had size $(4\pi/3)R_0^3$ and overdensity 
$\delta_0(R_0|R,z)$.  At $z$, it has size $R$.  
Repeat for an ensemble of such trajectories.  
So compute the distribution of first crossings of $B(R_0|R,z)$.  
The fraction of trajectories which first cross the barrier 
$B(R_0|R,z)$ at $R_0$ represents the fraction of mass that is in 
regions of mass $M_0$ that, at $z$, have size $R$.  
From this, the average number density $N(M_0|R,z)/V_{\rm tot}$ 
of such regions can be computed easily.  Assume that the number 
of such regions is conserved.  Then $N(M_0|R,z)V/V_{\rm tot}$ 
equals $p(M_0|R,z)$.  

\subsection{The halo distribution}\label{halod}
The previous subsection showed how to compute the Eulerian 
space distribution function using an excursion set model.  
However, the excursion set model can also be used to provide 
information about the halo distribution.  
This subsection shows why.

Following Bond et al. (1991), define a halo as a Lagrangian 
region that has collapsed to a vanishingly small Eulerian size: 
$R=0$.  Suppose $z$ is given, and consider the Eulerian scale 
$R=0$.  In the spherical collapse model, $B(R_0|0,z)$ is 
independent of $R_0$ but depends on $z$  (equation~\ref{mow}).  
Define $B(R_0|0,z)\equiv \delta_{\rm c}(z)$.  
As before, equate the fraction of trajectories 
$f(R_0,\delta_{\rm c})$ which first cross $\delta_{\rm c}(z)$ at 
$R_0$ with the fraction of mass in regions $M_0$ that, at $z$, 
have size $R=0$.  Following~(\ref{fsnm}), the average number 
density of such regions is $\bar n(M_0,\delta_{\rm c}) = 
N(M_0|0,z)/V_{\rm tot} = (\bar\rho/M_0)\,f(R_0,\delta_{\rm c})$, 
and $\bar n(M_0|z)$ is said to be the mass function of collapsed 
haloes.  This is exactly the excursion set model for the 
unconditional halo mass function developed by Bond et al. (1991).  
In the present context, the unconditional mass function is just the 
Eulerian distribution function in the limit $R=0$.  

Since the $R=0$ limit has this special interpretation, in what 
follows, it is convenient to make a distinction between the barrier 
crossing distribution in this limit, and that when $R>0$.  
Below the subscript `c' (for constant height), as in 
$f_{\rm c}(S_0,\delta_0)$, denotes the barrier crossing 
distribution when $R=0$, and the subscript `R' denotes the 
case $R>0$.  

Suppose we consider two barriers $B_1 = B(S_0|R_1,z_1)$ and 
$B_0 = B(S_0|R,z_0)$, with $z_1>z_0$.  (It may help to look at 
Fig.~\ref{barrier}.)  For now, assume that $R_1\le R$.  Let 
$f_{\rm R}(S_1,\delta_1|S_0,\delta_0)$ denote the fraction of 
trajectories which first cross $B_1$ at $S_1$ given that 
they first crossed $B_0$ at $S_0$.  Again, consider the limit
$R_1=R=0$ for both barriers.  Then 
$B_1 = \delta_{\rm c}(z_1) = \delta_{\rm c1}$, and 
$B_0 = \delta_{\rm c0}$ for all $S_0$, so both barriers 
have a constant height, and $B_1>B_0$.  
Bond et al. (1991) interpret $f_{\rm c}(S_1,\delta_1|S_0,\delta_0)$
as representing the fraction of mass in an $M_0$-halo that was 
earlier in $M_1$-haloes.  They define a conditional mass function 
as the average number of $M_1$-haloes that are within an $M_0$-halo:
${\cal N}(1|0) = (M_0/M_1)\,f_{\rm c}(1|0)$ if $M_1\le M_0$, 
and ${\cal N}(1|0)=0$ otherwise.  
In the context of this paper, $f_{\rm c}(1|0)$ represents the 
fraction of the Lagrangian region $V_0$ that has Eulerian size 
$R=0$ at $z_0$, which was previously in Lagrangian regions $V_1$ 
that, at $z_1$, also occupied vanishingly small Eulerian volumes.  
The reason for wording things in this way is that it shows how to 
compute other properties of the halo distribution in Eulerian 
space.  

For example, the average number of $M_1$ haloes that collapsed at 
$z_1$ and are in Eulerian regions $R>0$ at $z<z_1$ can be 
computed as follows.  Recall that $(M_1,z_1)$-haloes are associated 
with trajectories which cross the constant barrier 
$B_1=\delta_{\rm c1}$ at $S_1$.  
Since $\delta_{\rm c1}>\delta_{\rm c}(z)$, and since $B_0$ decreases 
monotonically as $S_0$ increases, each of these trajectories 
must have crossed $B_0=B(S_0|R,z)$ at some $S_0\le S_1$, so as to 
reach $\delta_{\rm c}(z_1)$ at $S_1$.  Therefore 
\begin{equation}
f_{\rm c}(S_1,\delta_{\rm c1}) = \int_0^{S_1} 
f_{\rm c}(S_1,\delta_{\rm c1}|S_0,\delta_{\rm 0})\,
f_{\rm R}(S_0,\delta_0)\ {\rm d}S_0 ,
\label{fseul}
\end{equation} 
where $\delta_0=B(S_0|R,z)$.  Since 
$f_{\rm c}(1) = (M_1/\bar\rho)\,\bar n(1)$, 
$f_{\rm c}(1|0) = (M_1/M_0)\,{\cal N}(1|0)$, 
$f_{\rm R}(0)\,{\rm d}S_0 = \Delta\,p(\Delta)\,{\rm d}\Delta$, 
and $\Delta=M_0/\bar\rho V$, equation~(\ref{fseul}) implies that 
\begin{equation}
\bar n(M_1,\delta_{\rm c1})V = \int_0^\infty \!\!\!
{\cal N}(M_1,\delta_{\rm c1}|M_0,\delta_0)\,
p(M_0|R,z)\ {\rm d}M_0 .
\label{nmav}
\end{equation}
If we set 
\begin{equation}
N(1|0) = {\cal N}(M_1,\delta_{\rm c1}|M_0,\delta_0),
\label{enns}
\end{equation}
where $N(1|0)$ was defined in Section~\ref{model}, just prior to 
equation~(\ref{nme}), and $\delta_0$ in ${\cal N}(1|0)$ is given 
by equation~(\ref{mow}), and we recall that 
$p(M_0)\,{\rm d}M_0 = p(\Delta)\,{\rm d}\Delta$ 
(equation~\ref{pdr}), then this expression is the same as 
equation~(\ref{nme}).  
Equation~(\ref{enns}) shows that the average number of $M_1$-haloes 
in Eulerian cells $V$ which contain mass $M_0$ at $z$ is equal to 
that in those Lagrangian regions $V_0$ which had initial density 
$\delta_0$ given by equation~(\ref{mow}).  

The main reason for writing this out explicitly is that it 
shows how statistics in the Lagrangian space can be used to 
compute statistics in Eulerian cells $V$.  Namely, the 
assumed conservation of number of regions in the two spaces 
(equation~\ref{nmav} equals equation~\ref{nme}) means that 
an average over Eulerian regions of size $V$ is equivalent to 
averaging over those Lagrangian regions that have evolved into 
Eulerian regions of size $V$ (equations~\ref{nmav} and~\ref{enns}).  
In practice, this means that Eulerian space quantities like 
$\bar\xi_{\rm hm}$ and $\bar\xi_{\rm hh}$ can be computed 
simply by using the appropriate value for $\delta_0$ (that 
given by equation~\ref{mow}) in the associated Lagrangian space 
quantities, and then summing over $M_0$, weighting the 
contribution from each $M_0$ appropriately.  This weighting is 
given by the barrier crossing algorithm, because the barrier 
crossing distribution $f_{\rm R}(S_0,\delta_0)$, with $\delta_0$ 
given by equation~(\ref{mow}), represents the fraction of Lagrangian 
space that is associated with regions of mass $M_0$ that, because 
they had initial overdensity $\delta_0$, are Eulerian regions of 
size $V$ at $z$.  

Section~\ref{linear} illustrates how this works.  It presents 
a model in which all these quantities can be computed analytically.  

\subsection{Relation to the work of Mo \& White}\label{mowhite}
Before moving on, it is interesting to compare this model 
with previous work.  Our equation~(\ref{enns}) is exactly the same 
as that assumed by Mo \& White (1996).  
They assumed that the Eulerian space quantity $N(1|0)$ could 
be computed by simply substituting the spherical collapse value 
of $\delta_0(R_0|R,z)$ into the Lagrangian formula ${\cal N}(1|0)$.  
They showed that the resulting formula for $N(1|0)$ provided good 
fits to the corresponding Eulerian space quantity measured in their 
numerical simulations of clustering. 
Since, in this regard, our model leads to the same formula as that 
of Mo \& White, the agreement with simulations provides some 
justification for the strong simplifying assumptions which lead to 
our model.  

Our model extends their results in the following way.  
Our equation~(\ref{nmav}) shows that, to compute averages in 
Eulerian space, knowledge of the Eulerian space distribution 
function $p(\Delta|R,z)$ is necessary to then take the correct 
average over Lagrangian regions $V_0$ that have become Eulerian 
regions $V$.  Mo \& White also knew this, but they did not know 
how to compute $p(\Delta|R,z)$.  Therefore, they assumed that it 
could be taken directly from their numerical simulations.  
However, as our approach shows, $p(\Delta|R,z)$ is related to 
the shape of the boundary (cf. our equation~\ref{fspd}).  
This relation must not be ignored.  
To see why, suppose (as Mo \& White did) that a Lognormal 
can be substituted for $p(\Delta|R,z)$.  If one uses the 
spherical model for $\delta_0(R_0|R,z)$ in ${\cal N}(1|0)$ and then 
does the integral on the right hand side of~(\ref{nmav}), one 
finds that it does not equal the correct value for the left hand 
side.  This is because, as our approach shows, self-consistency 
requires that $p(\Delta|R,z)$ depend on the boundary shape; 
once the boundary is specified, one is no longer 
free to choose any arbitrary distribution for $p(\Delta|R,z)$.  
This may partially explain why the Mo \& White results using a 
Lognormal are somewhat worse than when they use the distribution 
measured directly in their simulations.  

Since our approach shows how to derive $p(\Delta|R,z)$ from 
the boundary shape, our approach can be thought of as deriving, 
self-consistently, the Mo \& White (1996) spherical evolution 
formulae by a simple extension of the Bond et al. (1991) excursion 
set approach.  

\subsection{The two barrier problem}\label{twobare}
The previous subsection showed that problems involving the 
crossing of two barriers, in which one barrier was assumed 
to be constant, and the other not (i.e. the Eulerian scale 
was $R=0$ for one barrier, but $R>0$ for the other), could be 
used to provide information about the distribution of haloes 
within Eulerian cells $V$.  The case in which $R>0$ 
for both barriers is also interesting.  

Consider two barriers $B(R_0|R,Z)$ and $b(r_0|r,z)$.  
If $z=Z$ and $r<R$, then the two barrier problem is 
associated with the joint distribution of the mass within 
two concentric Eulerian cells.  Therefore, it may be useful
for estimating the evolved Eulerian space density profile.  
Of course, in the model, haloes collapse to zero radius, and 
this is unrealistic.  In practice, haloes virialize at some 
fraction of their turnaround radius.  Therefore, the solution 
of this two barrier problem is probably only useful on scales 
larger than the virial radius.  

If, instead, $r=R$, but $z>Z$, then the two barrier problem 
describes the matter within the same comoving cell at two 
different epochs.  That is, it can be used to quantify the way 
in which matter flows in and out of (comoving) Eulerian cells.  

\section{The linear barrier and white-noise initial conditions}\label{linear}
This section shows how the model described in Section~\ref{model} 
can be used.  It assumes that the initial Lagrangian space 
distribution is white-noise Gaussian and studies the first 
crossing distribution of a barrier whose height decreases linearly 
as $V_0=(4\pi/3)R_0^3$ decreases:
\begin{equation}
B(S_0|R,z) = \delta_{\rm c0}\,(1+z) - 
{\delta_{\rm c0}\,(1+z)\over\Delta},
\label{linbar}
\end{equation}
where $S_0 = 1/\bar\rho_0 V_0$, and $\Delta = M_0/\bar\rho V = (R_0/R)^3$.  

This barrier shape was chosen for a number of reasons.  
First, this shape is a simple approximation to the spherical 
collapse barrier (compare equation~\ref{mow}, and see 
Fig.~\ref{barrier}), and the associated dynamical evolution is 
similar.  Section~\ref{scoll} shows how the evolution of a spherical 
perturbation in this model differs from that in the usual 
spherical collapse model.  
Second, the first crossing distribution of a linear barrier is 
known, so the associated Eulerian distribution can be written 
analytically; it is Inverse Gaussian.  
The Inverse Gaussian is a good approximation to the distribution 
measured in numerical simulations of clustering from white noise 
initial conditions (Section~\ref{sims}), so the model may also be 
realistic.  Finally, for this barrier, the halo--mass and halo--halo 
correlations can all be computed analytically.  

\subsection{Relation to the spherical collapse model}\label{scoll}
The linear barrier studied in this section is associated with 
the following model for the collapse of objects.  

\begin{figure}
\centering
\mbox{\psfig{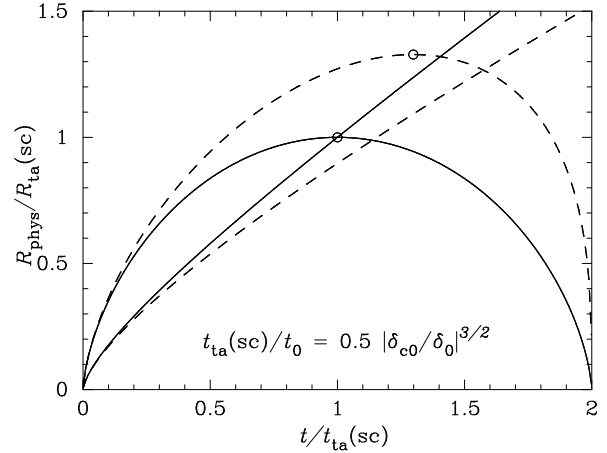}}
\caption{The physical radius of a perturbation in units of the 
spherical model turnaround radius, as a function of time in 
units of the spherical collapse model turnaround time.  
The solid curves show the spherical model, and dashed curves show 
the linear barrier  model studied in this paper.  
The two curves for each line type are for denser perturbations 
(which recollapse) and under-dense perturbations (which do not).}
\label{rphys}
\end{figure}

Let $R(z)$ denote the comoving size of an object at the epoch $z$.  
Then $R(z)=R_0$ initially.  If the object is in an underdense 
region, then its comoving size will increase, else it will 
decrease.  Trajectories with extrapolated linear overdensity 
$\delta_0$ greater than $\delta_{\rm c0}$ are associated with 
collapsed objects.  Collapsed objects have $R(z)=0$.  
Equation~(\ref{linbar}) implies that, before collapse, 
\begin{equation}
{R^3(z)\over R_0^3} = 1 - {\delta_0/(1+z)\over \delta_{\rm c0}}.
\end{equation}
The radius of an object in proper, physical coordinates 
is $R_{\rm p}(z) = R(z)/(1+z)$.  Objects which collapse have 
a turnaround radius---the maximum value that $R_{\rm p}(z)$ 
attains.  This occurs at 
\begin{equation}
(1+z_{\rm ta}) = {4\over 3}{\delta_0\over\delta_{\rm c0}} ,
\end{equation}
at which time 
\begin{equation}
{R(z_{\rm ta})\over R_0} = {1\over 4^{1/3}}.
\end{equation}

In comparison, collapsing objects in the spherical model reach 
turnaround at 
\begin{equation}
(1+z_{\rm ta}) = 4^{1/3}\,{\delta_0\over\delta_{\rm c0}},
\end{equation}
at which time 
\begin{equation}
{R(z_{\rm ta})\over R_0} = 
{(1+z_{\rm ta})\,R_{\rm p}(z_{\rm ta})\over R_0} = 
{6\over 10}{4^{1/3}\over \delta_{\rm c0}} = 
\left({4\over 3\pi}\right)^{2/3}
\end{equation}  
(e.g. Peebles 1980).  
Fig.~\ref{rphys} shows that, for overdense perturbations, 
turnaround in this model (dashed lines) occurs later, and at a 
larger radius, than in the spherical model (solid lines).  
To compensate, underdense regions expand less rapidly in 
this model than in the spherical model.  

\subsection{The constant barrier and statistics in Lagrangian space}
\label{const}
In the limit $R\to 0$, equation~(\ref{linbar}) for the 
barrier shape becomes independent of $S_0$, as it does for 
the spherical collapse model.  In this limit, the first 
crossing distribution is known (e.g. Bond et al. 1991), 
although we give a different derivation of it below.  
Therefore, to set notation, this section summarizes various known 
results about the crossing of a constant barrier by the 
Lagrangian trajectories described earlier, under the assumption 
that $\delta_0({\bmath r};0)$ is a Gaussian random field.  

Let $p(S_0,\delta_0)\,{\rm d}\delta_0$ denote the fraction of 
trajectories that have value between $\delta_0$ and 
$\delta_0 + {\rm d}\delta_0$ at $S_0$.  Also, let 
$p(S_0,\delta_0|S',\delta',{\rm first})\,{\rm d}\delta_0$ denote 
the fraction of those trajectories which first crossed the 
barrier $B(S_0|R,z)$ between $S'$ and $S'+{\rm d}S'$, 
which have value between $\delta_0$ and $\delta_0 + {\rm d}\delta_0$ 
at $S_0\ge S'$.  Then 
\begin{equation}
p(S_0,\delta_0) = \int_0^{S_0} 
p(S_0,\delta_0|S',\delta',{\rm first})\,f(S',\delta'|R,z)\ {\rm d}S' ,
\label{volt}
\end{equation}
provided that at $S_0$, $\delta_0\ge B(S_0|R,z)$.  
If both $p(S_0,\delta_0)$ 
and $p(S_0,\delta_0|S',\delta',{\rm first})$ are known, then 
this is a Volterra equation of the first kind, so it can be solved 
numerically, by forward substitution, to yield $f(S',\delta'|R,z)$.  

Suppose that the probability distribution of the density in 
the Lagrangian space is Gaussian.  Then 
\begin{equation}
p(S_0,\delta_0)\ {\rm d}\delta_0 = 
{{\rm d}\delta_0 \over\sqrt{2\pi S_0}}\ 
\exp\left(-{\delta_0^2\over 2S_0}\right) .
\label{pgaus}
\end{equation}
If the window $W$ is sharp in $k$-space, then 
\begin{equation}
p(S_0,\delta_0|S',\delta',{\rm first})\ {\rm d}\delta_0 = 
p(S_0,\delta_0|S',\delta')\ {\rm d}\delta_0 ,
\label{kspace}
\end{equation}
and 
\begin{equation}
p(S_0,\delta_0|S',\delta')\,{\rm d}\delta_0 = 
{ {\rm d}\delta_0 \over\sqrt{2\pi (S_0-S')}}
\exp\!\left[-{(\delta_0-\delta')^2\over 2(S_0-S')}\right]
\label{pjoint}
\end{equation}
(e.g. Bond et al. 1991).  
Thus, if the Lagrangian space distribution is Gaussian, and 
the window function is sharp in $k$-space, then the distribution 
of first crossing times is easy to compute, whatever the 
boundary shape.  Equations~(\ref{kspace}) and~(\ref{pjoint}) are 
also correct for a Top Hat filter in real space, if the Gaussian 
field is white noise.  So, the distribution of first crossing 
times is easy to compute for white noise also.  
In either of these cases, a trajectory $\delta_0(S_0)$ 
resembles the motion of a particle undergoing standard Brownian 
motion with zero drift.  Bond et al. (1991) and Lacey \& Cole (1993) 
used this fact to compute $f(S_0,\delta_0)$.  

Notice, however, that when the boundary is a constant, 
$B(S_0|R=0,z) = \delta_{\rm c0}(1+z) \equiv \delta_{\rm c}(z)$, 
then the form of $f(S_0,\delta_0)$ can be obtained directly 
from equation~(\ref{volt}).  First, take the 
derivative with respect to $\delta_0$ on both sides of 
equation~(\ref{volt}), and evaluate it at 
$\delta_0=\delta_{\rm c}(z)$.  Next, set 
$\delta'=\delta_{\rm c}(z)$.  
Then the integrand on the right hand side is zero, except when 
$S'=S_0$.  Thus, equation~(\ref{volt}) implies that 
\begin{eqnarray}
f_{\rm c}(S_0,\delta_{\rm c}[z]) &=& 
{\delta_{\rm c}(z)\over S_0}\,p(S_0,\delta_{\rm c}[z]) 
\nonumber \\
&=& {\delta_{\rm c}(z)\over \sqrt{2\pi S_0}}
\,\exp\left(-{\delta_{\rm c}^2(z)\over 2S_0}\right)
{{\rm dS_0}\over S_0},  
\label{fps}
\end{eqnarray}
where the subscript `c' denotes the fact that this is the 
distribution of first crossings of a constant boundary.  

If the trajectory is known to start from $(S_0,\delta_{\rm c0})$, 
rather than from the origin, then the distribution of 
first crossings of the constant barrier 
$\delta_{\rm c1}= \delta_{\rm c0}(1+z_1)$ 
can be solved similarly.  Provided 
$\delta_{\rm c1}>\delta_{\rm c0}$ and $S_1>S_0$, 
\begin{equation}
f_{\rm c}(S_1,\delta_{\rm c1}|S_0,\delta_{\rm c0}) = 
\left({\delta_{\rm c1}-\delta_{\rm c0}\over S_1-S_0}\right)
\,p(S_1,\delta_{\rm c1}|S_0,\delta_{\rm c0}).  
\label{fs1s0}
\end{equation}
These expressions for the first crossing distribution are the 
same as those derived by Bond et al. (1991) and 
Lacey \& Cole (1993).  

Following Bond et al. (1991), treat the parameter $\delta_{\rm c}(z)$ 
as a pseudo-time variable; it decreases as the universe evolves.  
Then associate the first crossing distribution of the barrier 
$\delta_{\rm c}(z)$ with a number density of regions containing 
mass $M_0$ in the Lagrangian space, and call every such region 
a dark matter halo.  Then the number density of $M_1$ haloes at 
the epoch labelled by $\delta_{\rm c1}=\delta_{\rm c}(z_1)$ is 
\begin{equation}
\bar n(M_1,\delta_{\rm c1})\ {\rm d}M_1 \equiv 
{\bar\rho_0\over M_1}\,f_{\rm c}(S_1,\delta_{\rm c1})\ {\rm d}S_1 .
\label{enm1d1}
\end{equation}
This is consistent with equation~(\ref{fsnm}), and 
$\bar n(M_1,\delta_{\rm c1})$ is sometimes called the unconstrained 
halo mass function at the epoch $z_1$.   

Similarly, the average number of $(M_1,\delta_{\rm c1})$-haloes 
that are within an $(M_0,\delta_{\rm c0})$-halo is 
\begin{equation}
{\cal N}(1|0) \equiv 
\left({M_0\over M_1}\right)\ 
f_{\rm c}(S_1,\delta_{\rm c1}|S_0,\delta_{\rm c0})\ 
{{\rm d}S_1\over {\rm d}M_1} 
\label{n10}
\end{equation}
(e.g. Lacey \& Cole 1993).  This is the constrained mass function.  

The joint distribution of the number $N_1$ of 
$(M_1,\delta_{\rm c1})$-haloes and the number $N_2$ of 
$(M_2,\delta_{\rm c1})$-haloes, that are both within the same 
$(M_0,\delta_{\rm c0})$-halo, averaged over all 
$(M_0,\delta_{\rm c0})$-haloes is 
\begin{equation}
C(12|0) \equiv 
\bigl\langle N_1N_2,\delta_{\rm c1}|M_0,\delta_{\rm c0}\bigr\rangle
= {\cal N}(1|0)\,{\cal N}(2|10) ,
\label{c120}
\end{equation}
where ${\cal N}(1|0)$ is given by equation~(\ref{n10}), and 
\begin{equation}
{\cal N}(2|10)\,{\rm d}M_2 = {M_0-M_1\over M_2}\,
f_{\rm c}(S_2,\delta_{\rm c1}|S_{01},\delta_{01})\ {\rm d}S_2 
\end{equation}
(Sheth 1996b).  
Here $S_{01} = S(M_0-M_1) = (M_0-M_1)^{-1}$ and $\delta_{01}$ is 
the overdensity in the remaining volume $V_0-V_1$ that is not 
occupied by the $M_1$-halo.  That is, 
\begin{equation}
1+\delta_{01} = {M_0-M_1\over \bar\rho_0(V_0-V_1)}.
\end{equation}
However, $M_0\equiv \bar\rho_0V_0(1+\delta_{\rm c0})$ and 
$M_1\equiv \bar\rho_0V_1(1+\delta_{\rm c1})$, so that, to 
lowest order,  
\begin{eqnarray}
\delta_{\rm c1} - \delta_{01} &=& 
(\delta_{\rm c1} - \delta_{\rm c0})\ 
{M_0(1+\delta_{\rm c1})\over M_0(1+\delta_{\rm c1})-M_1(1+\delta_{\rm c0})} 
\nonumber \\
&\approx& (\delta_{\rm c1} - \delta_{\rm c0})\ {M_0\over M_0-M_1} .
\end{eqnarray}  

Most of the formulae in this section are not new, but they will 
all be useful below.  
They have been included to set notation, and because, 
as the previous section showed, if these expressions are 
known, then simple transformations allow one to compute the 
associated Eulerian space quantities.  

\subsection{The linear barrier-crossing distribution and the 
matter distribution in Eulerian space}\label{pdfe}
The previous subsection described the first crossing distribution 
associated with a constant barrier by trajectories associated with 
a Lagrangian field that is Gaussian white noise.  Recall that 
the constant barrier is got from the linear barrier of 
equation~(\ref{linbar}) by setting $R=0$.  When $R>0$, then 
direct substitution shows that  
\begin{equation}
f_{\rm R}(S_0,\delta_0)\ {\rm d}S_0 = 
{B(0|R,z)\over\sqrt{2\pi S_0}}\ 
\exp\left(-{B^2(S_0|R,z)\over 2S_0}\right)\,{{\rm d}S_0\over S_0} ,
\label{invg}
\end{equation}
where $B$ is given by equation~(\ref{linbar}), 
satisfies equation~(\ref{volt}).  The identity 
\begin{equation}
\int_0^\infty \! f_{\rm R}(S_0,\delta_0)\ {\rm d}S_0 = 1,
\end{equation}
which follows from the definition of the modified Bessel 
function of the third kind, is useful in proving this result.  
A full derivation of this distribution is given in 
Appendix~\ref{deriv}.  

Equation~(\ref{invg}) is known as the Inverse Gaussian 
distribution.  It was first discovered in the context of 
Brownian motion by Schr\"odinger (1915).  Folks \& Chhikara (1978) 
give a review of its role in other fields.  Here, the distribution 
represents the fraction of Lagrangian space which is associated 
with regions of mass $M_0(S_0)$ provided the initial distribution 
is white-noise Gaussian, and the dynamics are a simple modification 
of the usual spherical collapse model (as described in the 
previous subsection.)  

For white noise, $M_0=1/S_0$, so, for white noise, 
equation~(\ref{invg}) in~(\ref{fspd}) implies that the 
Eulerian distribution function that is associated with the 
linear barrier is 
\begin{eqnarray}
p(\Delta|R)\ {\rm d}\Delta &=& 
\bar\rho V\,S_0\,f_{\rm R}(S_0,\delta_0)\ {\rm d}S_0\nonumber \\
&=& {1\over\sqrt{2\pi\bar\xi_{\rm m}\Delta}}\ \exp 
\left[-{(\Delta-\mu)^2\over 2\bar\xi_{\rm m}\Delta}\right]\ 
{{\rm d}\Delta\over\Delta} ,
\label{pdinvg}
\end{eqnarray}
where 
\begin{equation}
\bar\xi_{\rm m}(R) \equiv (\delta^2_{\rm c0}(1+z)^2\bar\rho V)^{-1} ,\ 
\ \ {\rm and}\ \ \mu = 1,
\label{params}
\end{equation}
and we have not bothered to write the $z$ dependence of 
$p(\Delta|R,z)$ explicitly.  
Notice that $p(\Delta|R)$ is also an Inverse Gaussian distribution.  
The mean of this distribution is $\mu=1$, and the variance is 
$\bar\xi_{\rm m}$.  The higher order moments of this distribution 
are simple.  If $\bar\xi_n$ is the $n$th cumulant of this 
distribution, then 
\begin{equation}
S_n\equiv {\bar\xi_n\over\bar\xi_{\rm m}^{n-1}} = (2n-3)!!
\label{esen}
\end{equation}
and it is independent of cell size $V$.  
This is simply a consequence of the fact that, in the model, 
haloes have vanishingly small Eulerian sizes:  $R=0$.  

As $\bar\xi_{\rm m}\to 0$, this Inverse Gaussian tends to the 
Gaussian distribution.  For fixed $V$, this happens if 
$\delta_{\rm c0}(1+z)\gg 1$.  Since $z$ is a pseudo-time variable, 
this means that, at sufficiently early times, the Eulerian 
distribution is Gaussian for almost all scales $V$.  This is 
sensible, since at early times, the Eulerian and Lagrangian 
distributions should be similar, and, by hypothesis, the Lagrangian 
distribution is Gaussian.  For a given $z$, $\bar\xi_{\rm m}\to 0$
as $V\gg 1$.  This means that, even at later times, the Eulerian 
distribution function, measured in sufficiently large cells $V$, 
appears Gaussian.  

When $V\to 0$, then the term in the exponent of 
equation~(\ref{pdinvg}) tends to 
$\Delta/\xi_{\rm m}=\delta^2_{\rm c0}(1+z)^2/S_0$, so 
$p(\Delta|R)\to V$ times equation~(\ref{enm1d1}), which is the 
same as $1/\Delta$ times equation~(\ref{fps}).  
This is consistent with the fact that, in this limit, 
equation~(\ref{linbar}) shows that the barrier shape tends to 
the constant $\delta_{\rm c0}(1+z)$, and the first crossing 
distribution of a constant barrier is given by equation~(\ref{fps}).  
This shows explicitly that the unconstrained mass function 
of~(\ref{enm1d1}) is essentially the same as the Eulerian 
distribution function, in the limit of vanishing cell size $V$.  
All these asymptotic relations are sensible.  

\subsection{The halo distribution}\label{halobar}
A little algebra shows that when the barrier shape is linear, 
then equation~(\ref{fseul}) is satisfied for all $R$.  
Recall that that expression simply expresses the fact that the 
fraction of trajectories which first cross the constant barrier 
$\delta_{\rm c1}$ at $S_1$ is equal to the fraction of trajectories 
which first cross the linear barrier (associated with the Eulerian 
scale $R$) at $\delta_0=B(S_0|R,z_0)$, and then cross the 
constant barrier $\delta_{\rm c1}\ge\delta_{\rm c0}$ at $S_1$, 
summed over all $S_0\le S_1$.  Of course, this implies that 
equation~(\ref{nmav}) is also satisfied:  
$\bar n(M_1,\delta_{\rm c1})V$ is equal to $V$ times 
equation~(\ref{enm1d1}), as it should.  
This shows explicitly that, for the linear barrier, the number 
density of haloes in the Eulerian space is equal to that in the 
Lagrangian space, and this number density is just what is required 
by the Bond et al. (1991) excursion set approach.  

The cross correlation between haloes and mass can be worked out 
similarly.  Start with equation~(\ref{xihme}).  
Use equation~(\ref{fspd}) to write $p(\Delta|R,z_{\rm obs})$ in terms 
of the Lagrangian barrier crossing distribution 
$f_{\rm R}(S_0,\delta_0)$.  Then use~(\ref{invg}) for 
$f_{\rm R}(S_0,\delta_0)$, and recall that 
$\delta_0=B(S_0|R,z_{\rm obs})$ where equation~(\ref{linbar}) 
gives $B$.  This means that $B(0|R,z)=\delta_{\rm c0}(1+z_{\rm obs})$, 
where $z_{\rm obs}\le z_1$.  Then use equation~(\ref{enns}) 
to set $N(1|0) = {\cal N}(1|0)$, and use~(\ref{n10}) for 
${\cal N}(1|0)$ with $\delta_0=B(S_0|R,z_{\rm obs})$.  
With these substitutions, the integral in equation~(\ref{xihme}) 
can be solved analytically:  
\begin{equation}
\bar\xi_{\rm hm}(1|R) = 
{M_1\over\bar\rho V}\left({1+z_1\over 1+z_{\rm obs}}\right) + 
{M_*(z_{\rm obs})\over \bar\rho V}
\left({z_1-z_{\rm obs}\over 1+z_1}\right),
\end{equation}
where 
\begin{displaymath}
\delta_{\rm c1} = \delta_{\rm c0}\,(1+z_1), \qquad 
\delta^2_{\rm c0}/S_0 = M_0/M_{*0}, 
\qquad {\rm and} 
\end{displaymath}
\begin{equation}
M_*(z_{\rm obs}) = M_{*0}\,(1+z_{\rm obs})^{-2}.
\label{dcz}
\end{equation}
Notice that when $z_1=z_{\rm obs}$, then 
$\bar\xi_{\rm hm} = M_1/\bar\rho V$.  That is, the only 
correlation that arises is that which is due to the fact that there 
is a halo of mass $M_1$ inside $V$, so there is certainly mass 
$M_1$ inside $V$.  
Since 
\begin{displaymath}
\bar\xi_{\rm m}(R) = 
{1\over\delta_{\rm c0}^2(1+z_{\rm obs})^2\bar\rho V} = 
{M_*(z_{\rm obs})\over \bar\rho V}, 
\end{displaymath}
equations~(\ref{params}) and~(\ref{dcz}) imply that 
\begin{equation}
{\bar\xi_{\rm hm}(1|R)\over\bar\xi_{\rm m}(R)} = 
{M_1\over M_*(z_{\rm obs})}\left({1+z_1\over 1+z_{\rm obs}}\right) + 
\left({z_1-z_{\rm obs}\over 1+z_1}\right) .
\label{xihmxim}
\end{equation}
This ratio is independent of the Eulerian cell size $V$.  
Again, this a consequence of the fact that the model assumes 
that haloes have zero volume, so if a halo is within a cell, 
then all its associated mass is also.  

This expression has an interesting relation to previous work.  
Mo \& White (1996) argue that, when 
$|\delta_{\rm c0}|\ll \delta_{\rm c1}$, then, in the limit 
of large $V$, 
\begin{equation}
\bigl\langle \delta_{\rm h}(1|0)\ \delta \bigr\rangle_{\rm R} \approx
\left(1 + {(\nu_1^2-1)/\delta_{\rm c0}\over \delta_{\rm c1}/[\delta_{\rm c0}(1+z_{\rm obs})]}\right)\, 
\bigl\langle\delta^2\bigr\rangle ,
\end{equation} 
where $\nu_1^2 \equiv \delta_{\rm c1}^2/S_1$.  Now, 
$\delta_{\rm c1}/\delta_{\rm c0} = (1+z_1)$, 
and $\langle\delta^2\rangle\approx\bar\xi_{\rm m}$ for large $V$, 
so this expression is the same as what is required by 
equations~(\ref{dcz}) and~(\ref{xihmxim}) if $\delta_{\rm c0}=1$.  
The extra factor of $\delta_{\rm c0}$ arises from the fact that, 
for large $V$ the density fluctuation $\delta$ in most Eulerian cells 
is small, so $\Delta\equiv 1+\delta\approx 1$ for most $V$.  
In this limit, equation~(\ref{mow}) shows that 
$\delta_0/(1+z)\approx\delta$ in the collapse spherical model, 
whereas equation~(\ref{linbar}) shows that it is 
$\approx\delta_{\rm c0}\,\delta$ for the linear barrier, 

Let $\bar\xi_{\rm hh}(M_1,M_2,\delta_{\rm c1}|R)$ denote the 
correlation function of $M_1$ and $M_2$ haloes identified 
at the epoch $z_1$, averaged over all comoving Eulerian 
cells $V$, at the epoch $z_{\rm obs}\le z_1$.  
Recall that this average can be computed in two steps.  All 
Eulerian cells can be classified by the mass $M_0$ within them.  
Further classify all Eulerian cells which contain mass $M_0$ 
by their Lagrangian overdensity $\delta_0$.  Since the number 
of such regions is conserved, the joint distribution of the 
number of $M_1$ and $M_2$ haloes, $N_1N_2$, averaged over all 
such $(M_0,\delta_0)$-regions, is the same in both spaces, 
provided we set $\delta_0=B(S_0|R,z)$.  
This average gives $C(12|0)$ of equation~(\ref{c120}), where 
$\delta_0=B(S_0|R,z)$ and $B$ is given by~(\ref{linbar}).  All that 
remains is to further average $C(12|0)$ over the number of such 
conserved regions.  This is done by weighting it by $p(\Delta|R)$, 
integrating over $\Delta$, and then dividing out the factors 
expected on average.  We have gone to the trouble of stating things 
this way to show that our definition of $\bar\xi_{\rm hh}(12|R)$ 
includes the effects of the scatter of halo counts in Eulerian 
regions which had the same overdensity as well as the same radius 
initially.  Therefore, $\bar\xi_{\rm hh}$ here denotes the 
same quantity as $\sigma^2_{\rm hh}$ of equation~(25) in 
Mo \& White (1996).  

Now, $1+\bar\xi_{\rm hh}(12|R)$ is given by equation~(\ref{xihhe}), 
where $C(12|0)$ is given by equation~(\ref{c120}).  
With equations~(\ref{fspd}) and~(\ref{invg}) for $p(\Delta|R)$, 
and setting $\delta_0=B(S_0|R,z)$ as before, this integral can be 
solved analytically.  The result is that 
\begin{equation}
{\bar\xi_{\rm hh}(12|R)\over \bar\xi_{\rm m}(R)} = 
{M_1 + M_2\over M_*(z_{\rm obs})}
\left({z_1-z_{\rm obs}\over 1+z_{\rm obs}}\right) + 
\left({z_1-z_{\rm obs}\over 1+z_1}\right)^2 ,
\label{xihhxim}
\end{equation}
where $z_1\ge z_{\rm obs}$, and $M_*(z_{\rm obs})$ is 
defined as before.  
Like the halo--mass correlation function, this ratio is 
independent of $V$ because, in the model, haloes are point sized.  
Thus, in the model, haloes are linearly-biased tracers of the 
mass on all scales.  Mo \& White (1996) argue that 
$\bar\xi_{\rm hh}/\bar\xi_{\rm m}$ should equal 
$(\bar\xi_{\rm hm}/\bar\xi_{\rm m})^2$ on large scales.  
Although our equation~(\ref{xihmxim}) for 
$\bar\xi_{\rm hm}/\bar\xi_{\rm m}$ is similar to theirs, 
the right hand side of our equation~(\ref{xihhxim}) is not the same 
as the square of the right hand side of equation~(\ref{xihmxim}); 
they differ by a factor of 
$[M(1+z_1)/M_*(z_{\rm obs})(1+z_{\rm obs})]^2$.  

When $z_1=z_{\rm obs}$, $\bar\xi_{\rm hh}(12|R)=0$, which implies 
that haloes, when they first virialize, are uncorrelated with 
each other.  Further, the correlation of haloes depends on the sum 
of the halo masses, but not on the masses themselves.  
This suggests that halo--halo correlations arise because of 
volume exclusion effects only.  That is, halo--halo correlations
arise from the fact that an $M_1$-halo occupies a region $V_1$ 
initially, so no other halo can occupy this region.  The 
correlation function is affected by the fact that a region is 
included, but not by the exact way in which the excluded region 
is populated.  

If the halo masses $M_1$ and $M_2$ and their formation epoch 
$z_1$ are fixed, then the halo--halo correlation function 
at a given comoving $R$ increases as $z_{\rm obs}$ decreases.  
This increase reflects the fact that haloes which formed at 
$z_1$, at which time they were uncorrelated with each other, 
must have merged with each other to construct the more massive 
haloes present at the later epoch $z_{\rm obs}\le z_1$.  
Suppose, instead, that the halo masses, the comoving scale, 
and the epoch at which the halo distribution is measured 
(i.e., $z_{\rm obs}$), are given, and we wish to consider the 
effect of changing the halo formation epoch $z_1\ge z_{\rm obs}$.  
Equation~(\ref{xihhxim}) shows that as $z_1$ decreases to 
$z_{\rm obs}$, $\bar\xi_{\rm hh}$ decreases.  That is, at fixed 
mass, the halo--halo correlation function in comoving coordinates 
exhibits negative evolution.  

\subsection{The two barrier problem}\label{twobar}
The previous subsections showed how the first crossing of a linear 
barrier by Brownian walks could be used to compute various 
interesting quantities associated with gravitational clustering.  
This section shows that the associated two barrier problem 
may also be useful.  
In general, the linear barrier has two free parameters 
which may be thought of as the $y$-intercept $\delta_{\rm c0}$ 
and the slope $\beta$, respectively:  
\begin{equation}
\delta_0(S_0) = \delta_{\rm c0} - \beta S_0.
\label{lindb}
\end{equation}
Below, we consider 
the statistics of trajectories which first cross a barrier with 
one choice of parameters, and then cross another.  

First, consider two barriers $\delta_1$ and $\delta_2$ which 
have the same $y$-intercept $\delta_{\rm c0}$, but different 
slopes, $\beta_1$ and $\beta_2$.  Assume that $\beta_2>\beta_1$.  
This means that trajectories must cross the barrier $\delta_2$ 
before they can cross $\delta_1$.  We seek an expression for the 
probability that a trajectory has its first crossing of 
$\delta_1$ at $S_1$, given that it has its first crossing of 
the barrier $\delta_2$ at $S_2\le S_1$.  
Recall the derivation of the barrier crossing distribution 
given in the Appendix.  To a particle that started from the origin 
and has just crossed the barrier $\delta_2(S_2)$, the barrier 
$\delta_1(S_1)$ has shape 
\begin{eqnarray}
\delta_{12}(S_1-S_2) &=& \delta_1(S_1)-\delta_2(S_2) \nonumber \\
&=& (\beta_2-\beta_1)\,S_2 - \beta_1(S_1-S_2)
\end{eqnarray}
Since this barrier is also linear, the solution is the same 
as before, except that $S_0\to S_1-S_2$ and 
$\delta_{\rm c}\to \delta_{12}$.  
Therefore, the first crossing distribution is  
\begin{displaymath}
f(S_1,\delta_1|S_2,\delta_2)\,{\rm d}S_1 = 
\end{displaymath}
\begin{equation}
{(\beta_2-\beta_1)S_2\over \sqrt{2\pi(S_1-S_2)}}
\exp\left(-{(\beta_2S_2 - \beta_1S_1)^2\over 2(S_1-S_2)}\right)
{{\rm d}S_1\over (S_1-S_2)} .
\end{equation}
Since $\delta_{\rm c0}$ is a pseudo-time variable, and it 
is the same for the two barriers, equation~(\ref{lindb}) shows 
that the expression above will be useful for computing statistics 
in concentric cells $V_1$ and $V_2>V_1$, at the same epoch.  
Therefore, it may be useful for computing Eulerian density 
profiles within the context of the model.  Since, in the model, 
haloes collapse to zero radius, this way of computing density 
profiles is probably only useful on scales larger than the virial 
radius.  The virial radius can be derived from combining the 
spherical collapse model with the fact that the mass within the
virial radius is given by the scale at which the trajectory first 
crosses $\delta_{\rm c0}$.  

The corresponding expression for two barriers which have the 
same slope $\beta$, but have different values for the 
$y$-intercept, say, $\delta_{\rm c1}$ and 
$\delta_{\rm c2}<\delta_{\rm c1}$ is 
\begin{displaymath}
f(S_1,\delta_1|S_2,\delta_2)\,{\rm d}S_1 = 
{(\delta_{\rm c1} - \delta_{\rm c2})\over \sqrt{2\pi(S_1-S_2)}}
\end{displaymath}
\begin{equation}
\ \ \ \times\ 
\exp\left(-{\bigl[(\delta_{\rm c1}-\delta_{\rm c2})-\beta(S_1-S_2)\bigr]^2\over 2(S_1-S_2)}\right){{\rm d}S_1\over (S_1-S_2)} .
\end{equation}
Equation~(\ref{lindb}) shows that this provides information 
about the matter in concentric cells $V_1$ and $V_2>V_1$, at 
two different epochs.  Since the value of $\beta$ is the same, 
the two cell sizes are related:  
$(V_2/V_1) = (\delta_{\rm c1}/\delta_{\rm c2})$.  

More interesting, perhaps, is the case in which the cell sizes 
$V$ are the same, but $\delta_{\rm c1}>\delta_{\rm c2}$.  This 
case is also more complicated, since now it is possible to 
cross barrier 1 before crossing barrier 2.  Provided 
$S_2<\bar\rho V$, 
\begin{displaymath}
f(S_1,\delta_1|S_2,\delta_2)\,{\rm d}S_1 = 
{\delta_{\rm c12}\over \sqrt{2\pi(S_1-S_2)}}
\end{displaymath}
\begin{equation}
\qquad \times\ 
\exp\left(-{\bigl[\delta_{\rm c12}-\beta_1(S_1-S_2)\bigr]^2\over 2(S_1-S_2)}\right){{\rm d}S_1\over (S_1-S_2)} ,
\end{equation}
where 
$\delta_{\rm c12} = (\delta_{\rm c1}-\delta_{\rm c2})(1-\beta_2)$,
and $\beta_1$ and $\beta_2$ are given by using~(\ref{linbar}) 
in~(\ref{lindb}), with $\delta_{\rm c1}$ and $\delta_{\rm c2}$, 
respectively.  
When the first crossing of $\delta_2$ occurs at $S_2>\bar\rho V$, 
then the trajectory will have crossed $\delta_1$ first.  In this 
case, $f(S_2,\delta_2|S_1,\delta_1)$ is given by the 
expression above, provided the labels 1 and 2 are interchanged.  
Bayes' rule then gives the associated 
$f(S_1,\delta_1|S_2,\delta_2)$, since
\begin{equation}
f(S_1,\delta_1|S_2,\delta_2) = 
f(S_2,\delta_2|S_1,\delta_1)
\,{f_{\rm R}(S_1,\delta_1)\over f_{\rm R}(S_2,\delta_2)} .
\end{equation}
These expressions contain information about the mass in the 
same comoving Eulerian cell at different epochs.  
They show that it is possible that the mass in 
$V$ first decreases and later increases.  This shows explicitly 
that, in the model, matter can flow in and out of Eulerian 
cells.  

\subsection{Comparison with simulations}\label{sims}
This subsection shows that the Inverse Gaussian distribution 
provides a good description of the evolved distribution for 
clustering from white noise initial conditions.  Before doing 
so, we first argue that such a test has, in fact, already been 
performed by others.  

\begin{figure*}
\centering
\mbox{\psfig{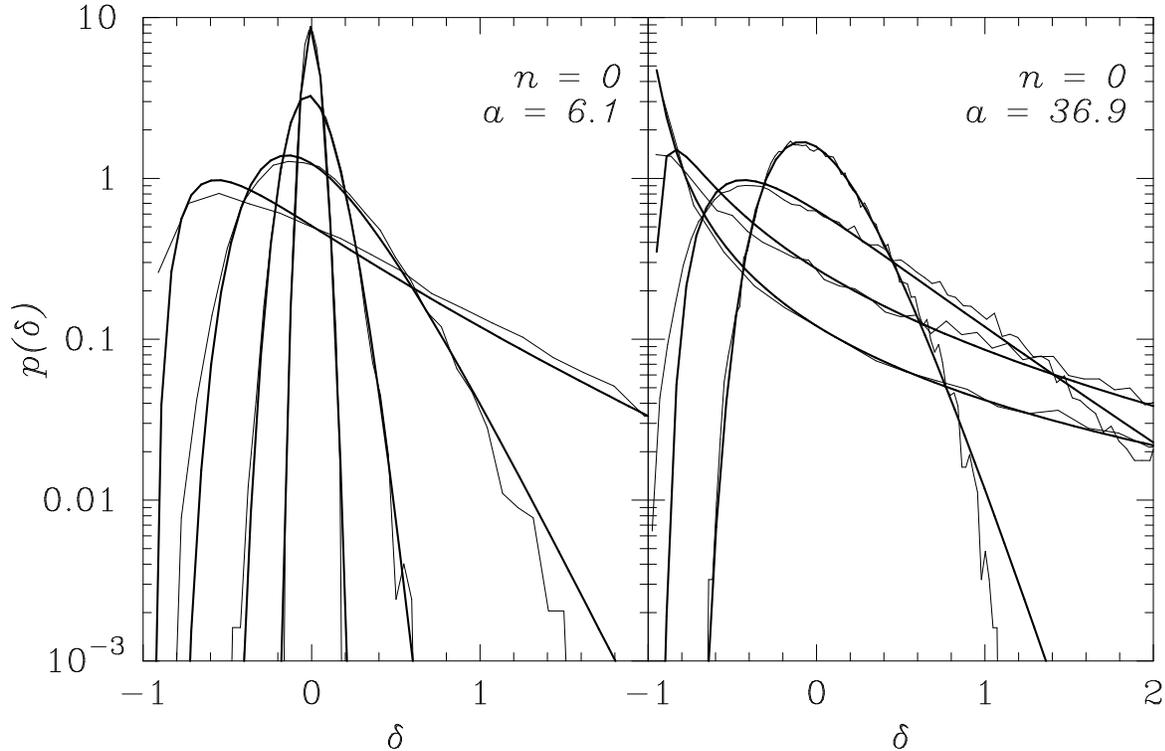}}
\caption{The Eulerian distribution function for a range of comoving 
scales, shown at two different output times, labelled by the 
expansion factor $a$ (initially $a=1$).  Thin curves show the 
distribution measured in simulations of clustering from white noise 
Gaussian initial conditions.  Thick curves show Inverse Gaussian 
distributions that have the same variance as the thin curves.  
Larger cells have narrower distribution functions.  }
\label{pdf0}
\end{figure*}

The Inverse Gaussian is a limiting form of the generalized Poisson 
distribution (GPD) 
\begin{equation}
p(N|R,b) = {\bar N_{\rm c}\over N!}\,
\bigl[\bar N_{\rm c} + Nb\bigr]^{N-1} \,
{\rm e}^{-\bar N_{\rm c} - Nb},
\label{gpd}
\end{equation}
where $N$ is the number of identical particles in a randomly 
placed cell of size $R$ which contains $\bar N$ particles on 
average, $\bar N_{\rm c}\equiv\bar N(1-b)$, and $0\le b\le 1$.  
In the astrophysical context, the GPD was first derived by 
Saslaw \& Hamilton (1984).  It can be related to the Inverse 
Gaussian as follows (Fry 1985; Sheth 1996a).  
The mean of this discrete distribution is $\bar N$, and the variance 
is $\bar N/(1-b)^2 \equiv \bar N(1+\bar N\bar\xi)$.  Notice that 
as $\bar N\to\infty$ and $b \equiv (1+\delta_{\rm c0})^{-1}\to 1$, 
$\bar N\bar\xi\to \delta_{\rm c0}^{-2}$, which is the relation 
required by equation~(\ref{params}).  In this limit, 
$\bar N\,p(N|R,b)$ tends to the Inverse Gaussian distribution 
$p(\Delta|R)$, since $N/\bar N \equiv \Delta$.  

Sheth (1998) shows that the GPD can be derived using  
a boundary crossing, excursion set model associated with 
the Poisson distribution in the initial Lagrangian space.  
The linear barrier shape considered in this paper is a limiting 
form of that associated with the Poisson case.  This correspondence 
is important, since the GPD is in good agreement with numerical 
simulations of clustering from Poisson initial conditions 
(Itoh, Inagaki \& Saslaw 1993 and references therein), provided 
the variance is allowed to be a free function of scale.  
Bouchet \& Hernquist (1992) showed that the GPD also describes 
clustering from white-noise initial conditions well.  
This just reflects the (intuitively obvious) fact that clustering 
from Poisson and white-noise initial conditions should evolve 
similarly.  Since the Inverse Gaussian is just a limiting form 
of the GPD (just as the white-noise Gaussian is a limiting 
form of the Poisson distribution), the Inverse Gaussian should 
provide a good approximation to the Eulerian distribution 
measured in simulations of clustering from white noise initial 
conditions, in the regime where discreteness effects are unimportant.  
Fig.~\ref{pdf0} shows this explicitly.  

The simulations of clustering from white noise initial conditions 
used here are the same as those studied by Mo \& White (1996), 
where they are described in more detail.  
The simulations contain $10^6$ particles in a cube with periodic 
boundary conditions.  
The figure shows the evolved Eulerian distribution function for 
a few representative choices of the comoving scale.  Each panel 
shows results for counts in spherical cells with comoving radii 
$0.02$, $0.04$, $0.08$, and $0.16$ times the box size.  
For each cell size, the thin curves show the distribution 
of counts averaged over $30,000$ cells.  
The two panels show results at two different expansion factors 
$a$, where $a=1$ initially.  

Comparison with the linear barrier model developed here is a little 
tricky.  In principle, the model requires that the free parameter 
$\bar\xi_{\rm m}$ of the Inverse Gaussian distribution 
(equation~\ref{pdinvg}) should scale as $\bar\xi_{\rm m}\propto 1/V$.
In fact, although $\bar\xi_{\rm m}$ in the simulations is 
$\propto 1/V$ initially, it scales differently at later times 
(e.g. Hamilton et al. 1991; Jain, Mo \& White 1995).  
Therefore, when comparing equation~(\ref{pdinvg}) with the 
simulations, on any scale $V$, we assume that $\bar\xi_{\rm m}(V)$ is 
given by the value measured in the simulations.  Since 
$\bar\xi_{\rm m}$ is related to the variance, the thick curves in 
Fig.~\ref{pdf0} show Inverse Gaussian distributions that have the 
same variance as the distributions shown by the thin curves.  
When this is done, the Inverse Gaussian (equation~\ref{pdinvg}) 
describes the simulation results reasonably well on all scales at 
all times.  In fact, the Inverse Gaussian fits $p(\delta)$ at least 
as well as the Press--Schechter mass function (equation~\ref{enm1d1}) 
fits the distribution of cluster masses (see, e.g., Lacey \& Cole 1994).  
Since the evolution of the variance is reasonably well understood 
(Hamilton et al. 1991; Jain, Mo \& White 1995) the results for the 
linear barrier presented here may be of more than just academic 
interest.  

For example, in the linear barrier model, when haloes 
first virialize, they are uncorrelated with each other 
(Section~\ref{halobar}).  What correlations do exist arise primarily 
from volume exclusion effects.  That is, in the initial Lagrangian 
space, haloes occupy a volume that is proportional to their mass.  
Since no other halo can occupy the region taken up by a given halo, 
this exclusion effect gives rise to (anti)-correlations, at least 
on small scales.  This is probably a generic consequence of the 
white-noise initial conditions, and not specific to the linear 
barrier model.  

\section{Discussion and extensions}\label{discuss}
\subsection{Summary of the model}
This paper develops a model that allows one to provide an 
approximate description of the spatial distribution of dark matter, 
as well as dark matter haloes, in a self-consistent way.  
It can also be used to estimate the density profiles of haloes, 
and to quantify how matter flows in and out of Eulerian cells.  
The model is described in Section~\ref{model}.  

The model assumes that, at the epoch $z$, the comoving Eulerian 
size $R$ of an initial Lagrangian region $R_0$, is related to 
its initial overdensity $\delta_0$.  This assumption is motivated 
by the spherical collapse model.  The additional assumption that 
the number of such regions is conserved, only their comoving size 
changes, allows one to compute statistics in the Eulerian space by 
taking appropriate averages over the relevant Lagrangian space 
quantities.  Since the initial Lagrangian distribution can be 
computed (e.g. Mo \& White 1996; Sheth \& Lemson 1998), so can 
the Eulerian distribution.  The algorithm for taking these 
appropriate averages is related to the solution of an excursion 
set, barrier crossing problem, in which the barrier shape is given 
by the spherical collapse model.  

A constant barrier is a special case of the barrier studied here.  
The first crossing distribution associated with a constant barrier 
can be used to estimate the distribution of virialized halo 
masses (Bond et al. 1991).  In the context of this paper, this 
special case corresponds to the limit of vanishing Eulerian cell 
size; the halo mass function is simply the Eulerian probability 
distribution function in the limit of vanishing Eulerian cell 
size.  Thus, the model here shows how the Bond et al. (1991) 
construction that yields the halo mass function can be extended 
to yield the Eulerian probability distribution function.  
The results of Mo \& White (1996) are also easily understood 
within the context of this model.  

In Section~\ref{linear}, the linear barrier was used to show explicitly 
how the model works.  This was done for a number  of reasons.  
Firstly, the spherical collapse barrier problem must 
be solved numerically, whereas, for the linear barrier, most 
interesting quantities can be computed analytically.  
Secondly, within the context of this model, the linear barrier 
can be understood as arising from a simple variant of the 
spherical collapse model (Section~\ref{scoll}).  
For this barrier shape, the associated Eulerian distribution 
is Inverse Gaussian (Section~\ref{pdfe}), and this is good 
agreement with numerical simulations of clustering from 
white noise initial conditions (Fig.~\ref{pdf0}).  
This provides yet another reason for considering the linear barrier.  

In the model, correlations between haloes and mass arise from the 
simple fact that a cell which is known to contain a halo certainly 
contains that halo's mass.  Correlations between haloes 
arise primarily from volume exclusion effects.  That is, in the 
initial Lagrangian space, haloes occupy a volume that is proportional 
to their mass.  Since no other halo can occupy the region taken up 
by a given halo, this exclusion effect gives rise to 
(anti)-correlations, at least on small scales.  This is probably a 
generic consequence of the white-noise initial conditions, and 
not specific to the linear barrier.  

For the linear barrier model, the evolution of halo--halo 
correlations can be written analytically (Section~\ref{halobar}).   
Section~\ref{twobar} showed that the two-linear-barrier 
problem can also be solved analytically; in the model, this can be 
used to provide information about the density run around random 
positions in Eulerian space.  Although the necessary formulae 
were all derived there, we did not pursue this further.  

\subsection{Scaling properties of the model}
The Inverse Gaussian distribution was derived here by considering 
the distribution of crossings of a linear barrier by Brownian 
motion.  The relation between this barrier shape 
(equation~\ref{linear}) and that required by the spherical collapse 
model (equation~\ref{mow}) was discussed in Section~\ref{scoll}.  

\begin{figure}
\centering
\mbox{\psfig{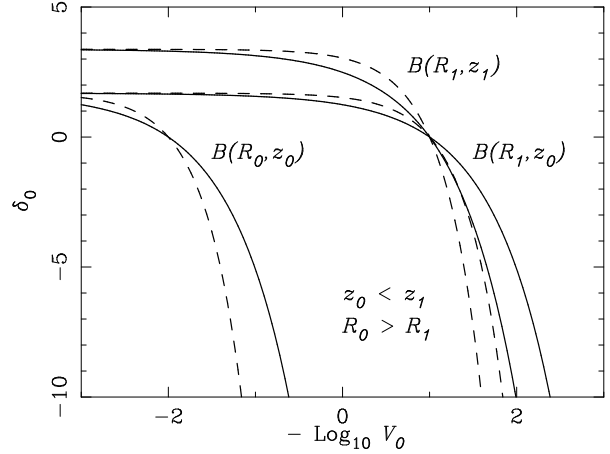}}
\caption{Dependence of the barrier shape on comoving Eulerian size 
$R$ and redshift $z$.  Solid curves show $B=\delta_0(R_0|R,z)$ 
of equation~(\ref{mow}), dashed curves show equation~(\ref{linbar}).
For white noise, $S_0\propto 1/V_0\propto 1/R_0^3$.}
\label{barrier}
\end{figure}

The thick solid lines in Fig.~\ref{barrier} show this spherical 
collapse barrier (equation~\ref{mow}) for representative values of 
$V\propto R^3$.  Curves for larger $R$ look just like those for 
smaller $R$ except that are shifted to the left (since the $x$-axis 
of the plot shows the negative of the logarithm of $V_0$).  
Curves for the same comoving $R$ but at higher redshift are simply 
multiplied by $(1+z)$, so the zero crossing remains the same.  
Dashed curves show the corresponding quantities for the linear 
barrier used in Section~\ref{linear}.  
This figure shows that changing the comoving size $R$ 
at fixed $z$ is equivalent to rescaling the $S_0$ axis, and changing 
$z$ at fixed comoving radius $R$ is equivalent to rescaling the 
$\delta_0$ axis.  This provides a strong constraint on the form of 
the distribution of first crossings.  

Namely, it implies that $f(S_0,\delta_0|R)$ should have a 
self-similar form, in that it should be a function of two parameters, 
one which is related to the scaling of the $S_0$ axis, and another 
which relates the scaling of the $S_0$ axis to that of the 
$\delta_0$ axis.  Since the first crossing distribution is a 
function of only two parameters, the associated Eulerian distribution 
is also.  It is convenient, then, to think of these parameters 
as being the mean and variance of the Eulerian distribution.  
This is an attractive feature of the model, since 
Colombi et al. (1997) find that, to a good approximation, the 
Eulerian distributions measured in $N$-body simulations of clustering 
from white noise initial conditions are functions of the mean and 
the variance only.  

\subsection{Scale--free initial conditions}
Initial Gaussian random fields with scale-free power spectra, 
$P(k)\propto k^n$ and $-3<n<1$, are a simple generalization 
of white-noise (for which $n=0$).  
Equation~(\ref{slin}) shows that, for these spectra, 
\begin{equation}
S_0\propto M_0^{-\alpha},\qquad {\rm where}\qquad \alpha = (n+3)/3,  
\label{s0n}
\end{equation}
which means that $\Delta = M_0/V = (S_{\rm V}/S_0)^{1/\alpha}$, 
where $S_{\rm V}\propto V^{-\alpha}$.  
This means that, when plotted as a function of $(\delta_0,S_0)$, 
the shape of the spherical collapse barrier depends on the initial 
power spectrum.  Although the barrier shape when $n\ne 0$ is 
different than for white noise, it still has the same scaling as 
the white noise barrier, since changing the Eulerian scale $R$ is 
still equivalent to a simple shift of the barrier along the 
$\log(S_0)$ axis.  So, the barrier crossing and the associated 
Eulerian distributions will still be functions of two parameters.   
This is in qualitative agreement with the results of 
Colombi et al. (1997).  

For initial conditions that are not scale-free, changing the 
Eulerian scale $R$ is no longer equivalent to a simple shift of 
the barrier along the $\log(S_0)$ axis.  Thus, the barrier 
crossing distribution is no longer self--similar, and 
the associated Eulerian distribution is no longer a function 
of just two parameters.  Of course, two parameter fits may 
remain a good approximation; at issue is the rate of change 
of the slope of the power spectrum over the scales at which 
most of the trajectories cross the barrier.  

While these are encouraging features of the model, extending it 
to describe clustering from initial conditions that are different 
from white noise is not straightforward.  The problem is one 
of ensuring correct normalization.  
For scale--free initial conditions, equations~(\ref{s0n})
and~(\ref{fspd}), and the requirement that the 
associated Eulerian distribution have unit mean 
(equation~\ref{norm}), means that the first crossing 
distributions must satisfy 
\begin{equation}
\int_0^\infty \!\!\! f(S_0,\delta_0|R)\,{\rm d}S_0 = 
\int_0^\infty \!\!\! \left({S_0\over S_{\rm V}}\right)^{1\over\alpha} 
f(S_0,\delta_0|R)\,{\rm d}S_0 = 1.
\label{fsalpha}
\end{equation}
The previous section showed that, for a white-noise Gaussian 
field, the distribution of first crossings of the linear barrier
has an associated Eulerian distribution which has unit mean, 
as required.  However, in general, not all barrier shapes are 
compatible with this requirement.  

For example, recall that equation~(\ref{bnrdo}) with 
$\delta_{\rm c0}=1.5$ is a good approximation to the spherical 
collapse model.  When $n=-1$, $\alpha = 2/3$, so the associated 
barrier, when written as a function of $S_0$, is the same (linear) 
as it was in Section~\ref{linear}.  
Now, for a linear barrier, the first passage distribution is Inverse 
Gaussian.  However, when $\alpha=2/3$ and $f(S_0,\delta_0|R)$ is 
Inverse Gaussian, then the second equality of equation~(\ref{fsalpha}) 
is not satisfied.  Thus, it appears likely that, for 
arbitrary power spectra, the excursion set approach developed 
here will not work if at least some parametrizations of the 
spherical collapse barrier are used.  

This means that it may be more fruitful to consider the 
inverse problem to that considered in this paper.  

\subsection{The inverse problem and the Generalized Inverse 
Gaussian distributions}
In this paper, the barrier shape was specified, then the first 
crossing distribution was obtained, and finally, the associated 
Eulerian distribution was derived.  It may be more useful to 
first specify an Eulerian distribution.  Then, transform it 
to a first crossing distribution, from which the shape of the 
barrier can be inferred.  This ensures that all normalization 
requirements have been satisfied.  Once the barrier shape is 
known, the halo--mass and halo--halo correlations can be worked 
out just as they were in this paper.  Finally, this barrier shape can 
be compared with that required by the spherical collapse model.  

In this context, it is worth mentioning that there is 
a class of distributions which provides convenient generalizations 
of the Inverse Gaussian, and may provide useful approximations to 
the Eulerian distributions measured in simulations of gravitational 
clustering from scale--free initial conditions.  

The Generalized Inverse Gaussian (GIG) distribution with index 
$\lambda$, scale parameter $\eta$, and concentration 
parameter $\omega$ is 
\begin{equation}
f_\lambda(x|\eta,\omega)\,{\rm d}x = 
{(x/\eta)^\lambda\over 2\,K_\lambda(\omega)}\,
{\rm e}^{-{\omega\over 2}({x\over\eta}+{\eta\over x})}\ 
{{\rm d}x\over x}, \qquad x\ge 0,
\end{equation}
where $K_\lambda(\omega)$ is a modified Bessel function of 
the third kind with index $\lambda$.  For a given $\lambda$, 
these are functions of just two parameters, $\eta$ and $\omega$, 
so they have the self--similar property discussed above.  
For these modified Bessel functions, 
\begin{eqnarray}
K_\lambda(\omega) &=& K_{-\lambda}(\omega) \qquad{\rm and} \nonumber \\
K_{\lambda+1}(\omega) &=& 2\,(\lambda/\omega)\,K_\lambda(\omega) 
+ K_{\lambda-1}(\omega).
\end{eqnarray}
Therefore, when $\lambda = -(2\alpha)^{-1}$, then the GIG 
distributions, with $x=S_0$, $\eta=S_{\rm V}$ and 
$\omega=\delta_{\rm c0}^2/S_{\rm V}$ satisfy 
equation~(\ref{fsalpha}), so the associated Eulerian distributions 
are correctly normalized and have unit mean.  When $n=0$, 
then $\alpha=1$, so $\lambda = -1/2$, and this distribution is the 
same as the Inverse Gaussian distribution considered in 
section~\ref{linear}.  

\begin{figure}
\centering
\mbox{\psfig{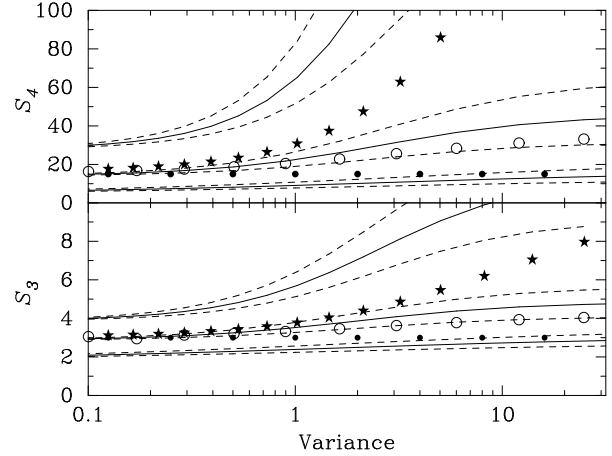}}
\caption{$S_3$ and $S_4$ versus variance for the Generalized 
Inverse Gaussian distributions associated with initial power 
spectra with slopes $n=0$ (filled circles), $-1$ (open circles), 
and $-2$ (stars).  
The solid curves show the values measured in numerical simulations 
of clustering from these initial conditions, as parametrized by 
Colombi et al. (1997).  Dashed curves show their estimates of 
the allowed range of values.}
\label{skew}
\end{figure}

The symbols in Fig.~\ref{skew} show the values of $S_3$ and $S_4$ 
(defined similarly to equation~\ref{esen}) plotted versus variance 
for these GIG distributions when $n=0$ (filled circles), 
$-1$ (open circles), and $-2$ (stars).  
The curves were computed using the extended perturbation theory 
fitting functions of Colombi et al. (1997).  
The solid curves show their fits to the values of $S_3$ and $S_4$ 
measured in numerical simulations of clustering from the 
corresponding scale-free initial conditions, and dashed curves 
give an estimate of the allowed range of values.  
While the GIG values do not fit the numerical results, they do 
show similar trends.  

The problem, then, is to determine the shape of the barrier for 
which the GIG distribution is the first crossing distribution, 
and to then compute the associated halo distribution.  
The solution of this inverse problem is the subject of ongoing work, 
where the barriers associated with the distributions given by the 
extended perturbation theory of Colombi et al. (1997) are also 
used as the Eulerian distributions.  

\section*{Acknowledgments}
I thank Houjon Mo and Simon White for providing the data from 
their simulations, and both of them, and Gerard Lemson, 
for interesting discussions.

\appendix
\section{The first passage time and the linear barrier}\label{deriv}

This Appendix presents a derivation of the first time $S$ 
that a particle undergoing Brownian motion with zero drift 
reaches the linear barrier 
\begin{equation}
\delta_\nu(S) = \nu - \beta\,S,
\end{equation}
having started from the origin $(S,\delta) = (0,0)$.  
Let $f_\nu(S)$ denote the first passage time distribution 
associated with this barrier.  The subscript $\nu$ indicates 
the height of the barrier at $S=0$.  

The assumption of zero drift means that, in the absence of the 
absorbing barrier, the mean distance of a particle from the $S$ 
axis, averaged over the ensemble of Brownian walks of the type 
shown in Fig.~\ref{bmotion}, is zero.  To be more precise, 
recall that equation~(\ref{pgaus}) gives the probability 
$p(S,\delta_0)\,{\rm d}\delta_0$ that a trajectory has value 
between $\delta_0$ and $\delta_0+{\rm d}\delta_0$, at $S$.  
So, for those trajectories, 
\begin{equation}
\langle\delta_0\rangle = 
\int \delta_0\,p(S,\delta_0)\ {\rm d}\delta_0 = 0 
\qquad\qquad{\rm for\ all}\ S,
\end{equation}
and 
\begin{equation}
\langle\delta_0^2\rangle = 
\int \delta_0^2\,p(S,\delta_0)\ {\rm d}\delta_0 = 
S + \langle\delta_0\rangle^2 = S\qquad{\rm at}\ S.
\end{equation}
Since $\langle\delta_0\rangle=0$ for all $S$, these are said 
to be Brownian walks with zero drift.  

Consider another linear barrier which is a simple shift of 
$\mu$ along the $\delta$ axis of the barrier above:  
\begin{equation}
\delta_\omega(S) = \mu + \delta_0(S) = 
\mu + \nu - \beta\,S \equiv \omega - \beta\,S .  
\end{equation}
Let $f_\omega(S)$ denote the first passage time distribution 
associated with this barrier.  The subscript $\omega$ indicates 
the height of the barrier when $S=0$.  

Then 
\begin{equation}
f_\omega(S)\,{\rm d}S = 
\int_0^S {\rm d}S'\,f_\nu(S')\,f_{\omega\nu}(S|S')\,{\rm d}S ,
\label{f1s}
\end{equation}
where $f_{\omega\nu}(S|S')$ denotes the first passage time 
distribution to $\delta_\omega(S)$, given that the particle 
started at $[S',\delta_\nu(S')]$ instead of the origin.  
Since the particle is undergoing Brownian motion, 
$f_{\omega\nu}(S|S')$ is the same as for a particle that 
starts at the origin, but sees a barrier 
\begin{equation}
\delta_\mu(S) = \delta_\omega(S) - \delta_\nu(S') = 
\mu - \beta\,(S-S').  
\end{equation}
So, we can write 
\begin{equation}
f_{\omega\nu}(S|S') = f_\mu(S-S'),
\end{equation}
where the subscript $\mu$ indicates that the height of the 
barrier above the starting position at $S'$ is $\mu$.  

This means that equation~(\ref{f1s}) is a convolution equation:
\begin{equation}
f_{\omega}(S) = f_{\mu+\nu}(S) = 
\int_0^S f_\nu(S')\,f_\mu(S-S')\ {\rm d}S' .
\label{conv}
\end{equation}
When $\beta = 0$, then the barrier is constant, and 
the first passage distribution is known to be 
\begin{equation}
f_\nu(S)\,{\rm d}S = \left({\nu^2\over 2\pi S}\right)^{1/2}
{\rm e}^{-\nu^2/2S}\ {{\rm d}S\over S} .
\end{equation}
It is straightforward to verify that this expression satisfies 
the convolution equation above.  Therefore, the solution to 
the convolution equation when $\beta\ne 0$ must be 
\begin{equation}
f_\nu(S)\,{\rm d}S = \left({\nu^2\over 2\pi S}\right)^{1/2}
{\rm e}^{-(\nu-\beta S)^2/2S}\ {{\rm d}S\over S} .
\end{equation}
A more formal derivation can be found in, e.g., 
Schr\"odinger (1915) or Cox \& Miller (1967).  
The following derivation follows that in Kao (1996) closely.

First, multiply both sides of 
equation~(\ref{f1s}) by $\exp(-tS)$, and integrate over all $S$.  
This gives the Laplace transform ${\cal L}(\omega,t)$ of 
$f_\omega(S)$.  Equation~(\ref{f1s}) implies that 
\begin{eqnarray}
{\cal L}(\omega,t) &=& 
\int_0^\infty {\rm d}S\ {\rm e}^{-tS}
\int_0^S {\rm d}S'\ f_\nu(S')\,f_{\omega\nu}(S|S') \nonumber \\
&=& \int_0^\infty {\rm d}S'\ f_\nu(S') 
\int_{S'}^\infty {\rm d}S\ f_\mu(S-S')\ {\rm e}^{-tS} \nonumber \\
&=& \int_0^\infty {\rm d}S'\ f_\nu(S')\,{\rm e}^{-tS'}
\int_0^\infty {\rm d}S''\ f_\mu(S'')\,{\rm e}^{-tS''} \nonumber \\
&=& {\cal L}(\nu,t)\ {\cal L}(\mu,t),
\end{eqnarray}
which is the same as 
\begin{equation}
{\cal L}(\nu+\mu,t) = {\cal L}(\nu,t)\ {\cal L}(\mu,t).
\label{laplas}
\end{equation}
This is a functional equation with solution 
${\cal L}(\nu)={\rm e}^{-C\nu}$, where $C>0$ is some constant, 
and we have not bothered to write the dummy variable $t$.  The  
problem, then, is to find $C$.  

To do so, notice that the crossing of a barrier which decreases as 
$S$ increases, by Brownian walks with zero drift, is equivalent 
to the crossing of a constant barrier by trajectories that, in 
the mean, drift upwards as $S$ increases, provided that the drift 
is chosen correctly.  This correct choice of drift 
corresponds to choosing 
\begin{equation}
\langle\delta_0\rangle \equiv \beta\,S 
\qquad\qquad\qquad{\rm at}\ S.
\label{drift1}
\end{equation}
Of course, the variance remains the same, so that 
\begin{equation}
\langle\delta_0^2\rangle =  
S + \langle\delta_0\rangle^2 = S + \beta^2 S^2
\qquad\qquad{\rm at}\ S.
\label{drift2}
\end{equation}
So the problem of the crossing of the linear barrier 
$\delta_\nu(S) = \nu - \beta S$ by walks with zero drift is 
transformed to the problem of the crossing of a constant barrier 
of height $\nu$ by walks with upward drift $\beta$.  In particular, 
the relation~(\ref{laplas}) remains the same.  

So, to find $C$, assume that the probability of first crossing the 
barrier $\nu$ in the first small increment $s$ along the $S$ axis 
is negligible.  Suppose that at $S=s$, the random 
trajectory has value $\delta_0<\nu$.  Then 
\begin{eqnarray}
f_\nu(S) &=& 
\int f_\nu(S|\delta_0,s)\,p(s,\delta_0)\ {\rm d}\delta_0 \nonumber \\
&=& \int f_{\nu-\delta_0}(S-s)\,p(s,\delta_0)\ {\rm d}\delta_0 ,
\end{eqnarray}
so that 
\begin{eqnarray}
{\cal L}(\nu,t) &=& \int_0^\infty {\rm d}S\,{\rm e}^{-tS} 
\int f_{\nu-\delta_0}(S-s)\,p(s,\delta_0)\ {\rm d}\delta_0 \nonumber \\
&=& \int {\rm d}\delta_0\ p(s,\delta_0)
\int_s^\infty {\rm d}S\ f_{\nu-\delta_0}(S-s)\,{\rm e}^{-tS} \nonumber \\
&=& {\rm e}^{-ts}
\int {\rm d}\delta_0\ p(s,\delta_0)\ {\cal L}(\nu-\delta_0,t) .
\end{eqnarray}
In what follows, we will sometimes omit the dummy variable $t$:  
${\cal L}(\nu)\equiv{\cal L}(\nu,t)$.  As $s$ and $\delta_0$ 
are small, 
\begin{equation}
{\cal L}(\nu-\delta_0)\approx {\cal L}(\nu) - 
\delta_0\,{\cal L}'(\nu) + {\delta_0^2\over 2}{\cal L}''(\nu),
\end{equation}
and 
\begin{equation}
{\rm e}^{-ts} \approx 1 - ts.
\end{equation}
These expressions imply that 
\begin{eqnarray}
{\cal L}(\nu) &\approx& (1-ts)\,
\left({\cal L}(\nu)-\langle\delta_0\rangle\,{\cal L}'(\nu)+{\langle\delta_0^2\rangle\over 2}\,{\cal L}''(\nu) \right)\nonumber\\
&=& (1-ts)
\left({\cal L}(\nu) - \beta s\,{\cal L}'(\nu)+{s + (\beta s)^2\over
2}\,{\cal L}''(\nu) \right)\nonumber \\
&\approx& {\cal L}(\nu) - ts\,{\cal L}(\nu)
- \beta s\,{\cal L}'(\nu) + {s\over 2}\,{\cal L}''(\nu),
\end{eqnarray}
The second line follows from~(\ref{drift1}) and~(\ref{drift2}), 
and the third is correct to lowest order in $s$.  
Dividing by $s$, and taking the limit $s\to 0$, reduces this to 
\begin{equation}
t\,{\cal L}(\nu) = -\beta\,{\cal L}'(\nu)
+ {1\over 2}\,{\cal L}''(\nu).
\label{diffeq}
\end{equation}
However, 
\begin{displaymath}
{\cal L}(\nu) = {\rm e}^{-C\nu},\ \ {\rm so}\ 
{\cal L}'(\nu) = -C{\rm e}^{-C\nu},\ \ {\rm and}\ 
{\cal L}''(\nu) = C^2{\rm e}^{-C\nu}. 
\end{displaymath}
These expressions in~(\ref{diffeq}) imply that 
\begin{equation}
C^2 + 2 \beta C - 2t = 0,
\end{equation}
so that, if $C>0$, then 
\begin{equation}
C = -\beta + \sqrt{\beta^2 + 2t},
\end{equation}
since $\beta$ and $t$ are both positive.  Thus, 
\begin{equation}
{\cal L}(\nu,t) = {\rm e}^{\nu\beta - \nu\sqrt{\beta^2 + 2t}}.
\end{equation}
Inverting this Laplace transform gives the Inverse Gaussian 
distribution.  

This problem can also be formulated in the context of the 
diffusion equation.  In the notation of Lacey \& Cole (1993), 
the associated diffusion equation is 
\begin{equation}
{\partial Q\over\partial S} = 
-\beta\, {\partial Q\over\partial\delta}
+ {1\over 2}{\partial^2 Q\over\partial\,\delta^2},
\end{equation}
where $\beta$ is the drift term, and $Q(S,\delta,\nu)$ represents 
the solution to this equation in the presence of an absorbing 
boundary at $\delta=\nu$.  Bond et al. (1991) and 
Lacey \& Cole (1993) considered the case $\beta=0$.  

\end{document}